\documentclass{ar2e}
\renewcommand\baselinestretch{1}
\begin{document}



\input epsf.sty

\input psfig.sty

\newcommand{\nwat}{\mbox{\textrm{nW m$^{-2}$ sr$^{-1}$}}}
\newcommand{\lesssim}{\ensuremath{< \atop \sim}}
\newcommand{\grtsim}{\ensuremath{> \atop \sim}}
\newcommand\apj{$\it Ap. J. $ }
\newcommand\aj{$\it Astron. J. $ }
\newcommand\aap{$\it Astron. Astrophys. $ }
\newcommand\mnras{$\it MNRAS $ }
\newcommand\araa{$\it Ann. Rev. Astron. Astrophys. $ }
\newcommand\pasp{$\it Publ. Astron. Soc. Pac. $ }
\newcommand\apjs{$\it Ap. J. Suppl. $ }
\newcommand\nature{$\it Nature $ }
\setcounter{tocdepth}{3}    

\jname{Annu. Rev. Astron. Astrophys.}
\jyear{2000}
\jvol{1}
\ARinfo{1056-8700/97/0610-00}

\title{THE COSMIC INFRARED BACKGROUND: MEASUREMENTS AND
 IMPLICATIONS}  

\markboth{HAUSER \& DWEK}{COSMIC INFRARED BACKGROUND}

\author{Michael G. Hauser
\affiliation{Space Telescope Science Institute, 3700 San Martin Drive,
Baltimore, Maryland 21218, e-mail: hauser@stsci.edu}
Eli Dwek \affiliation{Laboratory for Astronomy and Solar Physics, Code 685,
NASA/Goddard Space Flight Center, Greenbelt, Maryland 20771, e-mail: 
eli.dwek@gsfc.nasa.gov}}


\begin{abstract}
The cosmic infrared background records much of the radiant energy released by processes of structure formation that have occurred since the decoupling of matter and radiation following the Big Bang.  In the past few years, data from the Cosmic Background Explorer ({\it COBE}) mission provided the first measurements of this background, with additional constraints coming from studies of the attenuation of TeV
$\gamma$-rays.  At the same time there has been rapid progress in resolving a significant fraction of this background with the deep galaxy counts at infrared wavelengths from the Infrared Space Observatory ({\it ISO}) instruments and at submillimeter wavelengths from the Submillimeter Common User Bolometer Array (SCUBA) instrument.  This article reviews the measurements of the infrared background and sources contributing to it, and discusses the implications for past and present cosmic processes.  
\end{abstract}

\maketitle

\section{INTRODUCTION}
\label{intro}
One of the outstanding challenges in modern cosmology is to explain the formation
of structure in the universe.  The assembly of matter into stars and galaxies and the subsequent evolution of such systems is accompanied by the release of radiant energy powered by gravitational and nuclear processes.  Cosmic expansion and the absorption of short wavelength radiation by dust and re-emission at long wavelengths will shift a significant part of this radiant energy into infrared background radiation.  A cosmic infrared radiation background is therefore an expected relic of structure formation processes, and its measurement provides new insight into those processes.  Until a few years ago, there had been no definite measurements of the infrared background radiation.

\begin{figure}
\epsfxsize=4.75in	
\epsfbox{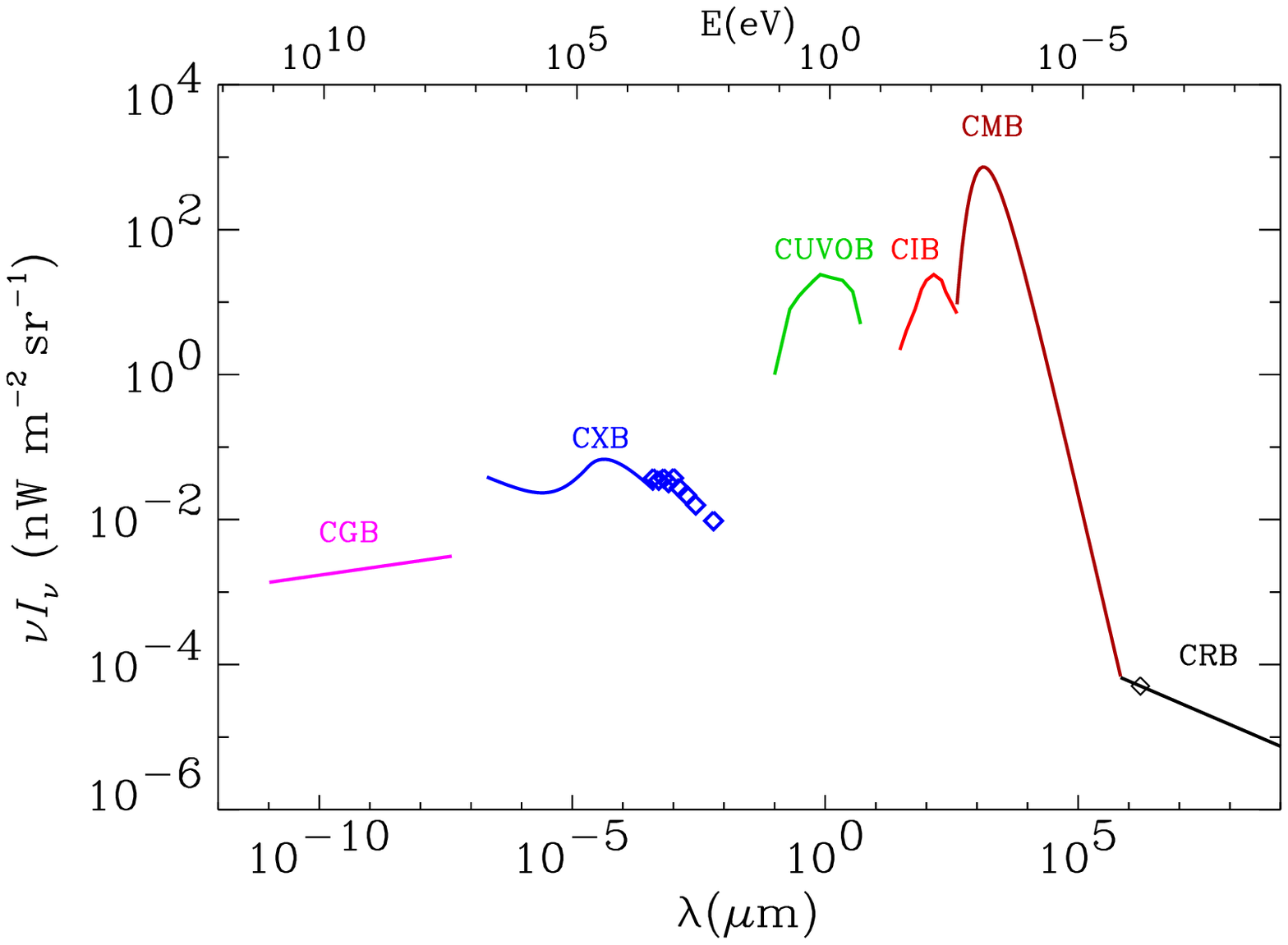}
\caption{}
\label{glob_ebl}
\end{figure}

For perspective, we show in Figure~\ref{glob_ebl} the
spectrum of the extragalactic background radiation over $\sim$20 decades of energy, 
from radio waves (10$^{-7}$ eV) to high energy $\gamma$-ray
photons of a few hundred GeV.  The cosmic microwave background (CMB), the fossil blackbody radiation from the
Big Bang, is the dominant form of electromagnetic energy. Figure~\ref{glob_ebl} gives only a schematic representation of the spectrum at wavelengths from the ultraviolet (UV) to the far infrared, roughly based on the energy released in producing the heavy elements.  
The extragalactic background light from UV 
to far infrared wavelengths, which we shall refer to as the EBL, is seen 
likely to be the dominant radiant energy in the universe aside from the CMB.  The background light in the more limited spectral range from 1--1000~$\mu$m, excluding the CMB, will be referred to as the cosmic infrared background (CIB).  As we discuss in this review, most radiant backgrounds shown in Figure~\ref{glob_ebl} other than the CMB are 
causally connected.  


In spite of the recognized significance of the CIB, its measurement has remained elusive because of the bright foreground radiations from which it must be distinguished.  The observational evidence has changed dramatically in the past few years, with the first direct measurements of this radiation provided by NASA's {\it Cosmic Background Explorer} ({\it COBE}) satellite, and additional evidence coming from the Japanese {\it Infrared Telescope in Space} ({\it IRTS}).  Indirect evidence constraining the CIB is coming from the rapidly developing ability to measure intergalactic attenuation of $\gamma$-rays at TeV energies.  Coincidentally, there have been rapid advances in ground and space observations which are resolving at least some of the sources of the CIB.  Since this field is now very dynamic, the emphasis of this review will be on the observational advances since the mid-90's and their implications.  Other summaries of both observational and theoretical work may be found in papers presented at numerous conferences and references therein (Lawrence 1988; Bowyer \& Leinert 1990; Holt et al 1991; Rocca-Volmerange et al 1991; Longair 1995; Calzetti et al 1995; Kafatos \& Kondo 1996; Dwek 1996; Lemke et al 2000; Franceschini 2001; Harwit \& Hauser 2001).  We treat sources within the solar system and Milky Way galaxy as undesired foregrounds to be discriminated from the CIB, and do not describe them in any detail (see Leinert et al 1998 for a recent compilation). 

The plan of this review is as follows.  In $\S$~\ref{history} we sketch the history of the growing interest in the infrared background radiation.  Section~\ref{cib_obs} provides a description of the observational evidence for the CIB and a summary of the UV-optical background.  In $\S$~\ref{impl} we address direct implications of the measurements.  Section~\ref{models} deals with how the measurements constrain evolutionary models.  In $\S$~\ref{sum_fut} we summarize the main conclusions and address future prospects.

We uniformly present photometric results in terms of $\nu I_{\nu}$, where $I_{\nu}$ is the spectral intensity at frequency $\nu$.  A unit of convenient size is the \nwat.  Conversion to $I_{\nu}$ in MJy\,sr$^{-1}$ (1 MJy\,$\equiv 1$ megajansky = $10^{-20}$\,W\,m$^{-2}\,$Hz$^{-1}$)
can be done with the relation  
\begin{displaymath}
\nu I_{\nu} (\nwat~) = [3000/\lambda (\mu {\rm m)}] I_{\nu} {\rm(MJy\, sr}^{-1}).
\end{displaymath}
Conversion to energy density, $\varepsilon ^2\, n_{\varepsilon}$, is given by 
\begin{displaymath}
\varepsilon ^2 n_{\varepsilon} {\rm (eV\, cm}^{-3}) = 2.62\, \times\, 10^{-4}\ \nu I_{\nu} (\nwat~), 
\end{displaymath}
where $\varepsilon\ $ is the photon energy in eV and  $n_{\varepsilon}$ is the photon spectral number density in photons cm$^{-3}\,$eV$^{-1}$.  Throughout this review we express the Hubble constant, $\mathrm{H_0}$,  as \mbox{$\mathrm{H_0}$ = 100\,$h$ km\,sec$^{-1}\,$Mpc$^{-1}$}.

\section{HISTORICAL REVIEW}
\label{history}
Recognition that the brightness of the night sky is an important astronomical datum dates back at least to Olbers (1826; see Harrison 1990 for an historical review).  Early calculations of the optical background radiation due to galaxies within the context of general relativity were carried out by Shakeshaft (1954), McVittie \& Wyatt (1959), Sandage \& Tammann (1964), and Whitrow \& Yallop (1964; 1965).  Not surprisingly, these authors dealt primarily with integrated starlight.  Whitrow \& Yallop (1965) also took account of starlight absorption by intervening galaxies and by intergalactic dust.  Re-emission of energy at infrared wavelengths by the absorbing dust was not considered.

The discovery of the cosmic microwave background (Penzias \& Wilson 1965) provided 
strong support for the notion of a hot, evolving early
universe, and led to recognition of the significance of a 
cosmic infrared background for cosmology.  In such a 
universe, one would expect a CIB, distinct from the CMB, associated with the
formation of structure and condensation of luminous
objects from primordial neutral matter following the decoupling 
of matter and radiation at a redshift $z$~$\sim$~1100.  
Peebles considered this infrared background in 1965 (unpublished lecture; see Peebles (1993), pp. 146--147), and noted the lack of direct knowledge 
of the sky brightness in the three decades of wavelength from 
1 to $1000~\mu$m.  The only observational limit on the CIB, which suggested that it was not large enough to close the universe, was provided by the presence of 10$^{19}$ eV cosmic ray protons, which would have been attenuated by photo-pion production by such a large CIB (Peebles 1969).

Partridge \& Peebles (1967a) recognized that young galaxies must have been 
more luminous than more evolved systems in order to produce the present metal abundances, and studied the possibility of detecting them individually for various epochs of formation.  In a second paper, Partridge \& Peebles (1967b) also calculated the integrated infrared background that would be produced by their model young galaxies in various cosmological scenarios. The effect of dust was ignored, so the background was generally brightest in the 
\mbox{1--10~$\mu$m} range. They compared these predictions with estimates of the 
foreground radiation from solar system and Galactic sources, and correctly concluded that 
the CIB is much fainter than the foregrounds.  Their work was important in 
stimulating observational programs to measure the CIB.

Harwit (1970) reviewed the status of early attempts to measure the infrared sky brightness.  He noted the importance of background measurements for understanding discrete classes of objects, such as QSOs, which had been found to be highly luminous in the far infrared (Kleinmann \& Low 1970; Low 1970).  
Measurement of the far-infrared background would set limits on the number and duration of such luminous episodes.  The high far-infrared luminosity of galaxies in the local universe had led Low \& Tucker (1968) to predict an infrared background peaking at a wavelength longer than 50~$\mu$m with a total energy of order 1--10\% of that in the recently discovered CMB.  As we show (\S~\ref{total_energy}), this prediction was substantially correct.  Harwit  also reviewed the relationship between low-energy photon backgrounds and energetic cosmic ray electrons, protons, and $\gamma$-rays.  Through the processes of inverse Compton scattering, photo-pion production, and electron-positron pair production respectively, the CIB provides a source of opacity to each of these cosmic radiations.  As of the time of Harwit's review, only rather high upper limits had been obtained on the CIB.  Even the bright diffuse foreground radiations were poorly known due to the difficulty of such measurements.  Longair \& Sunyaev (1972) presented a similarly uncertain description of the infrared background in a comprehensive review of extragalactic background radiation across the entire electromagnetic spectrum.  

The early theoretical estimates of the infrared background ignored the effect of partial thermalization of starlight by dust.  Since the mid-'70s, many investigators have taken this into account in models of varying degrees of sophistication and complexity, e.g., Kaufman (1976), Stecker et al (1977), Negroponte (1986), Bond et al (1986, 1991), Hacking \& Soifer (1991), Beichman \& Helou (1991), and Franceschini et al (1991, 1994).  The modeling efforts listed here were all carried out before the CIB had been detected.  Present knowledge of the CIB provides important new constraints on models for its origin (\S~\ref{models}).  

\section{OBSERVATIONS}
\label{cib_obs}

\subsection{Measurement Challenges}
\label{meas_chall}
The CIB has few defining observational characteristics on which to base a detection.  The radiation is of extragalactic origin, and is therefore expected to be isotropic on large scales.  There is no distinctive spectral signature.  The spectrum will depend in a complex way on the characteristics of the
luminosity sources, on their cosmic history, and on the history of dust formation and the distribution of dust relative to the luminosity sources.  Because discrete sources contribute at least part of the CIB, the background will have 
fluctuations superimposed on the isotropic signal. In $\S$~\ref{meas_chall}  to $\S$~\ref{irts} we review direct measurements of the CIB based on searches for isotropic, extragalactic infrared radiation.  Studies of the fluctuations are discussed in $\S$~\ref{fluc_anal}. 

Direct measurement of the infrared background is both technically and astrophysically very challenging.  The technical challenge is to make absolute sky brightness measurements relative to a well-established zero flux level.  Emission from telescope and instrument components and the Earth's atmosphere must be eliminated.  Scattered and diffracted light from the very bright local sources (Sun, Earth, and Moon) must also be strongly rejected.  In practice, this requires that observations be conducted with carefully designed, cryogenically cooled instruments located above the Earth's atmosphere.  Confident measurement of the CIB requires sufficient observation time to identify and eliminate potential sources of systematic measurement errors.  

\begin{figure}
\epsfxsize=4.75in	
\epsfbox{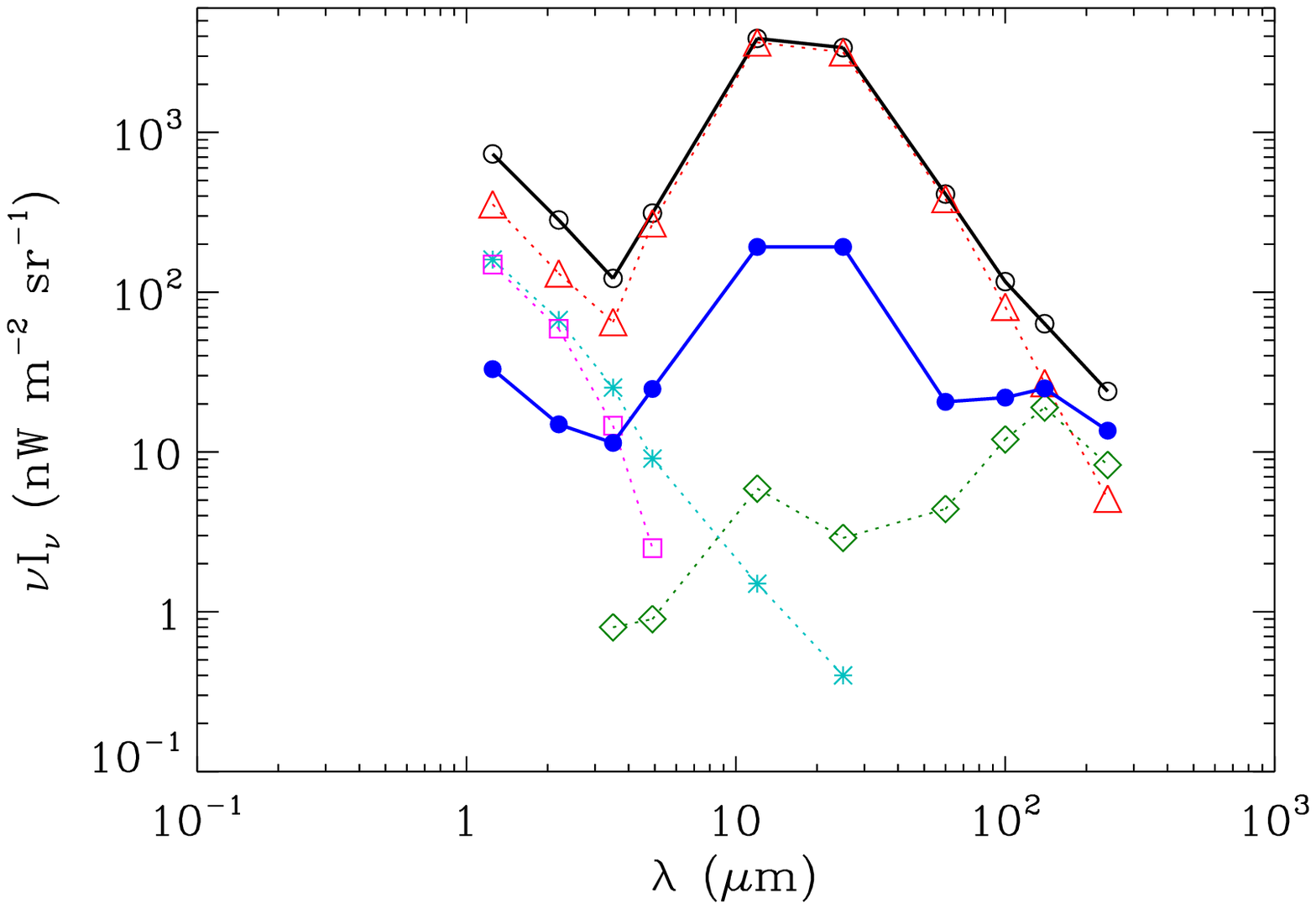}
\caption{}
\label{lh_plot}
\end{figure}

The fundamental astrophysical challenge for direct CIB measurement is definitive discrimination of the CIB from the many bright celestial contributors to the sky brightness.  These include discrete sources, such as stars and other compact sources within the Galaxy, and diffuse sources such as light scattered and emitted by interplanetary dust (IPD) and emitted by interstellar dust.  At wavelengths longer than $\sim$400~$\mu$m, the CMB   becomes dominant and must be discriminated from the CIB.  Figure~\ref{lh_plot} shows the measured spectral energy distribution of the sky in the direction of minimum H~I column density in the Galaxy, the 'Lockman Hole' 
at Galactic coordinates $(l,b)\sim(150^\circ,+53^\circ)$ [geocentric ecliptic coordinates 
$(\lambda,\beta)\sim(137^\circ,+45^\circ)]$ (Lockman et al 1986).  Figure~\ref{lh_plot} also shows the individual contributions from the foreground sources as determined by the COBE team (Section~\ref{dirbe_meas}). Even at 
this high ecliptic and Galactic latitude, the largest contribution to the sky brightness from 1.25 to 140~$\mu$m comes from the interplanetary dust. Starlight is substantial from 1.25 to 3.5~$\mu$m, and interstellar dust emission is strong for wavelengths greater than 60~$\mu$m.  There are two spectral windows most favorable for finding a faint extragalactic background: (1) the near infrared window near 3.5~$\mu$m, which is the minimum between scattered and emitted light from the IPD, and (2) the submillimeter window between $\sim$100~$\mu$m, the peak of the interstellar dust emission, and the CMB.


After discrimination and removal of the foregrounds, 
a candidate detection requires a positive residual signal significantly in excess of the random and systematic uncertainties associated with the measurements and foreground discrimination processes.  This residual signal must be isotropic, and must not plausibly be associated with any contribution from the solar system or Galaxy.  Few measurements of the infrared sky brightness have conclusively demonstrated all  three necessary conditions for detection of the CIB, i.e.,  that the signal is  significantly positive, that it is of extragalactic origin, and that it is isotropic.

\subsection{Background Measurements from Rockets}
\label{rockets}
The early attempts to observe the CIB were mostly carried out with instruments on sounding rockets.  Rocket-borne measurements of the diffuse infrared sky brightness continued into the 1990's from near to far infrared wavelengths (Murdock \& Price 1985; Matsumoto et al 1988a, 1988b; Noda et al 1992; Matsuura et al 1994; Kawada et al 1994). 
The rocket-based efforts increased our understanding of the dominant astrophysical contributions to the infrared sky brightness, and provided some upper limits or claims of possible detection of the CIB.  However, the limited observation time and sky coverage, and the systematic error issues associated with rocket observations, such as emissions from rocket exhaust, residual atmosphere, and Earthshine, prevented achieving compelling evidence for detection of the CIB.

\subsection{Background Measurements from {\it IRAS}}
\label{iras}
The {\it Infrared Astronomical Satellite} ({\it IRAS}) carried out the first all-sky survey in the infrared, mapping the sky at 12, 25, 60 and 100~$\mu$m (Neugebauer et al 1984).  The {\it IRAS} survey instrument measured the total sky brightness, though it was designed primarily for detection of discrete sources.  It did not contain a cold shutter or other internal means to establish or monitor the instrumental zero point for the brightness measurements.  Nevertheless, an approximate absolute brightness scale was determined (Beichman et al 1988).  The {\it IRAS} data clearly revealed large scale, diffuse emission components of the sky brightness from the solar system and Galaxy (Hauser et al 1984; Low et al 1984; reviewed by Beichman 1987).  Though several authors noted potential evidence for an extragalactic background component in the {\it IRAS }100~$\mu$m data (Rowan-Robinson 1986;  Rowan-Robinson et al 1990; Boulanger \& P\'erault 1988), the uncertainties in the {\it IRAS} zero point calibration and in the removal of foreground contributions precluded firm detection of the CIB in these studies.

\subsection{Background Measurements from {\it COBE}}
\label{cobe_cib}

The strongest evidence for direct detection of the CIB to date comes from studies based on the {\it COBE} measurements.  For this reason we describe these measurements in some detail.  

\subsubsection{THE {\it COBE} INSTRUMENTS}
\label{cobe_insts}
The {\it COBE} mission carried two instruments designed to make absolute sky brightness measurements (Boggess et al 1992).  The Diffuse Infrared Background Experiment (DIRBE) was designed primarily to search for the CIB from 1.25 to 240~$\mu$m.  The Far Infrared Absolute Spectrophotometer (FIRAS) was designed primarily to make a definitive measurement of the spectrum of the CMB, and to extend the search for the CIB from 125~$\mu$m to millimeter wavelengths.  

The {\it COBE} instruments, spacecraft, orbit, sky scan strategy and data processing were all designed to optimize the ability to make these difficult measurements with minimal stray light contamination and systematic measurement uncertainties.  Data collection at cryogenic temperatures extended over ten months, providing ample time for careful calibration and tests for systematic errors such as photometric offsets and stray light contamination.

The DIRBE instrument (Silverberg et al 1993) was an absolute photometer which provided maps of the full sky in 10 broad photometric bands at 1.25, 2.2, 3.5, 4.9, 12, 25, 60, 100, 140, and 240~$\mu$m.  A summary of the DIRBE investigation and the initial results of its CIB measurements is provided by Hauser et al (1998).  Additional information is given in the {\it COBE} DIRBE Explanatory Supplement (Hauser et al 1997).   The DIRBE was designed for extremely strong stray light rejection for any source, discrete or diffuse, out of the field of view of the instrument.  The stray light was demonstrated to be less than 1~\nwat~at all wavelengths.  The instrument contained a cold chopper, which provided continual measurement of the sky brightness relative to that of a cold internal beam stop.  It had a full beam cold shutter, which was closed frequently to measure the instrumental radiative and electronic zero point offsets.  The offsets were stable, with uncertainties less than a few~\nwat~in all bands.
The instantaneous field of view was 0.7$^\circ$ x 0.7$^\circ$, a compromise between ability to discriminate stars and ability to map the whole sky with high redundancy every six months.  Redundancy allowed monitoring of the apparent annual variations in sky brightness due to the Earth's orbital motion within the IPD cloud.  This variation is a unique signature of the IPD contribution to the sky brightness.  The sensitivity  (1$\sigma$) of the instrument per field of view averaged over the ten months of cryogenic operation was $\sim$2~\nwat~at 1.25 and 2.2~$\mu$m, 0.5--1~\nwat~from 3.5 to 100~$\mu$m, and  \mbox{33 (11) \nwat} at 140 (240)~$\mu$m.  Repeated observations of stable celestial sources provided photometric closure over the sky, and assured reproducible photometry to $\sim$1\% or better for the duration of the mission.  Calibration of the DIRBE flux scale was based on observations of a few isolated infrared sources of known brightness.

The FIRAS instrument was a Fourier transform spectrometer in the form of a polarizing Michelson interferometer.  The instrument was operated in a null, differential mode, providing precise spectral comparison of the sky brightness with that of a very accurate full beam blackbody calibrator at wavelengths from 100~$\mu$m to 1~cm.  The instrument and calibration are discussed by Mather et al (1993), Fixsen et al (1994), Mather et al (1999), and Brodd et al (1997).  The FIRAS had a 7$^\circ$ diameter field of view, with extremely low sidelobe response over a broad spectral band.  The sensitivity (1$\sigma$) of the instrument from 500~$\mu$m to 3~mm was 0.8~\nwat~ per field of view averaged over the ten months of cryogenic operation.  The photometric calibration errors  (1$\sigma$) associated with the precision blackbody calibrator were typically 0.02~MJy\,sr$^{-1}$, corresponding to $\nu I_\nu \sim$ 0.3 and 0.1~\nwat~at 200 and 600~$\mu$m wavelength respectively.  Fixsen et al (1996) and Mather et al (1999) showed that the rms deviation of the CMB spectrum from that of a $2.725\pm 0.002$ K blackbody was less than 50 parts per million of the peak CMB brightness from 476~$\mu$m to 5~mm wavelength.  

Fixsen et al (1997) compared the DIRBE and FIRAS zero point and absolute gain
calibrations in their range of spectral overlap.  They found that the calibrations of the two instruments were consistent within the quoted uncertainties of each and the systematic 
uncertainties of the comparison.  The comparison is relatively weak for the DIRBE 100~$\mu$m band due to the low FIRAS sensitivity there and partial coverage of the DIRBE bandpass.  Since the FIRAS systematic calibration uncertainties are smaller than those of the DIRBE, these results can be used to make small systematic corrections in the DIRBE 140 and 240~$\mu$m results, though DIRBE-based results in the literature use the DIRBE calibration.  

The {\it COBE} data have provided direct detections of the CIB in the near-infrared and submillimeter windows, and limits elsewhere within the spectral range of the two instruments.  These results, described in \mbox{$\S$~\ref{dirbe_meas}} and \mbox{$\S$~\ref{firas_meas}}, are listed in Table~\ref{cib_tab} and are shown in Figure~\ref{cib_plot}.  Searches for fluctuations in the CIB and their implications for the CIB,  discussed in \S~\ref{fluc_anal}, are listed in Table~\ref{fluc_tab} and are also shown in Figure~\ref{cib_plot}.



\subsubsection{CIB DETECTIONS AND LIMITS BASED ON DIRBE DATA}
\label{dirbe_meas}
The first tentative identification of the CIB using DIRBE data was reported by Schlegel et al (1998).  In the course of preparing Galactic reddening maps based on the far infrared data from {\it IRAS} and DIRBE, they removed the zodiacal emission contribution from the DIRBE data using the DIRBE 25~$\mu$m map as a template.  Correlating the residual maps with H~I maps they found a significant constant background at 140 and 240~$\mu$m which they identified as a possible measurement of the CIB.  However, they were not certain that the residual constant could not be an instrumental effect. They could not therefore be confident that the CIB had been detected, and noted that the results of the DIRBE team would be more definitive.  

The DIRBE team searched for the CIB in all 10 DIRBE wavelength bands (Hauser et al 1998).  Using the final photometric reduction of the DIRBE data, they reported conservative upper limits at all wavelengths based upon the darkest measured sky brightness at each wavelength.  These dark sky limits (95\%~CL) are $\nu I_\nu~<~$398, 151, 64, 193, 2778, 2820, 315, 94, 73, and 28~\nwat~at 1.25, 2.2., 3.5, 4.9, 12, 25, 60, 100, 140 and 240~$\mu$m respectively.  After modeling and removing the contributions from foreground sources, they found positive residuals exceeding 3$\sigma$ only at 4.9, 100, 140 and 240~$\mu$m, of which only the 140 and 240 $\mu$m were found to be isotropic (Table~\ref{cib_tab}).  Isotropy tests included (1) tests of the consistency of the mean residuals on five sky patches near the two Galactic poles, the two ecliptic poles, and in the Lockman Hole; (2) tests for spatial variations such as gradients with Galactic and ecliptic latitude; (3) tests of the residual pixel brightness distributions; (4) tests for correlations of the residuals with the foreground models; and most rigorously (5) a two-point correlation function analysis of the residuals on 2\% of the sky including areas at high north and south Galactic latitudes.  Only the 140~$\mu$m and 240~$\mu$m residuals were isotropic in the correlation function test, though due to detector noise the limits on anisotropies at these wavelengths were only of the order of 10\% of the mean CIB value.  

The DIRBE team showed that the 140 and 240~$\mu$m residuals could not be explained in terms of unmodeled components of the solar system or Galaxy (Dwek et al 1998), and therefore concluded that the CIB had been measured at these wavelengths.  At all other wavelengths, they reported CIB upper limits based upon the residuals and estimated uncertainties.  A conservative lower limit of 5~\nwat~at 100~$\mu$m was also determined, based on the argument that a thermal source producing the spectrum detected at 140 and 240~$\mu$m could not be fainter than this level at 100~$\mu$m (Dwek et al 1998).
The DIRBE team reported the 140 and 240 $\mu$m background both on the DIRBE photometric scale and transformed to the more accurate FIRAS photometric scale ($\S$~\ref{cobe_insts}).  The recalibrated brightnesses  are consistent with the values on the DIRBE scale, the most notable difference being about a 1.5$\sigma$ reduction at 140~$\mu$m (Table~\ref{cib_tab}).

Systematic uncertainties in the foreground determination dominated the uncertainties reported by the DIRBE team.  The interplanetary dust contribution presents the most difficult problem except at the longest wavelengths (Figure~\ref{lh_plot}).  The IPD brightness was determined by fitting the apparent annual variation of the sky brightness over the whole sky with a parameterized model of the spatial distribution and scattering and radiative properties of the dust cloud (Kelsall et al 1998).  Though this model was quite successful, reproducing the seasonal variation well and leaving evident map artifacts only at the few percent level, the uncertainties in the small differences between the measured brightness and the model are substantial.  Furthermore, it is not possible to determine a unique IPD model from measurements of the sky brightness and its apparent annual variation made from within the cloud.  Kelsall et al considered a number of different shapes for the cloud density distribution that gave comparably good fits to the time dependence.  They estimated the model uncertainty from the spread between these models in residual brightness at high Galactic latitude.  

Arendt et al (1998) described the DIRBE team methods for determining the Galactic contributions to the sky brightness.    Discrete sources brighter than the DIRBE direct detection limit ($\sim$ 5th magnitude at 2.2~$\mu$m) and regions of bright extended emission were blanked out of the DIRBE maps.  The contribution from fainter discrete Galactic sources was calculated from 1.25 to 25~$\mu$m using a faint source model based on the detailed statistical SKY model of Wainscoat et al (1992).    

The contribution from the interstellar medium (ISM) at most DIRBE wavelengths was identified by its variation over the sky.  The 100~$\mu$m map was used to construct an ISM template for all other wavelengths from 12 to 240 $\mu$m.  The residual maps with IPD and discrete Galactic sources removed at each wavelength was correlated with the ISM template to determine a global ISM color relative to 100~$\mu$m for each map.  The ISM template was then scaled by each color and subtracted from the corresponding map to produce the final residual maps.  The residual maps from 12 to 60 $\mu$m failed the isotropy tests, and their mean residuals at high Galactic latitude provided upper limits on the CIB.

The emission at 100, 140 and 240~$\mu$m (after removal of the IPD contribution) was correlated with H~I column density in two well-studied dark fields near the North ecliptic pole and the Lockman Hole. The correlation line was then extrapolated to zero H~I column density.  If all of the Galactic infrared emission were correlated with H~I gas, extrapolation of this correlation to zero column density would yield the extragalactic light.  If there were additional infrared emission associated with other gas components not correlated with H~I, this procedure would not yield accurate results.  Arendt et al (1998) chose these two fields because they were known to have little molecular gas, and low limits on the contribution from ionized gas regions could be set using pulsar dispersion measures and H$\alpha$ measurements.
These additional determinations were averaged with the mean map residuals at 100, 140 and 240~$\mu$m to obtain the final CIB values shown in Table~\ref{cib_tab}.

To test the 100~$\mu$m residual for isotropy, an ISM template independent of the 100~$\mu$m map was needed.  Arendt et al used the H~I map of Stark et al (1992) as an ISM spatial template for this purpose.  The final 100~$\mu$m residual map did not pass the isotropy tests, so only an upper limit on the CIB was reported.  There was no discernible ISM contribution at 1.25 and 2.2~$\mu$m.  Arendt et al determined the small ISM contributions at 3.5 and 4.9~$\mu$m using an analysis based on a reddening-free parameter.  The uncertainty in the ISM contribution at these short wavelengths made a negligible contribution to the overall uncertainty.

Following publication of the DIRBE team results, there have been numerous analyses aimed at improving the determination of the foregrounds.  Recognizing that the bright IPD foreground is a major obstacle to CIB detection, Finkbeiner et al (2000) analyzed the DIRBE data using two approaches which avoid the need for a detailed IPD model.  Both methods were based on analysis of the annual variation of dimensionless ratios of the data along opposing lines-of-sight after removal of the ISM contribution, and both showed the presence of seasonally varying contributions from interplanetary dust plus a temporally constant background at 60 and 100~$\mu$m.  Their analysis required assumptions about the spatial symmetries and temporal invariance of the IPD cloud similar to those of the Kelsall et al (1998) model.    Finkbeiner et al tentatively identified the constant backgrounds with the CIB (Table~\ref{cib_tab}), though they could not conclusively demonstrate an extragalactic origin.  They further noted that the values are rather high and not consistent with limits implied by the opacity of the intergalactic medium to TeV $\gamma$-rays  ($\S$~\ref{tev_lims} and \ref{gamma_opac}).  They concluded that there is not yet a satisfactory explanation for these constant backgrounds.

The statistical model of the stellar contribution used by the DIRBE team was a major source of uncertainty in the near infrared.   There have been several studies aimed at reducing that uncertainty.  Dwek \& Arendt (1998) created a full sky template of stellar light directly from the DIRBE data by assuming that the CIB at 2.2~$\mu$m is equal to the integrated light from galaxies (7.4 \nwat~at the time of their analysis).  They constructed a map of starlight at 2.2 $\mu$m by subtracting this assumed CIB value and the zodiacal light model of Kelsall et al (1998) from the DIRBE 2.2~$\mu$m map.  This residual 2.2~$\mu$m map was used as a starlight template at 1.25, 3.5 and 4.9~$\mu$m.  They obtained a significantly positive residual only at 3.5~$\mu$m.  They tentatively identified this 3.5~$\mu$m residual as the CIB, though they did not demonstrate that it was isotropic.  Table~\ref{cib_tab} shows updated 1.25, 3.5, and 4.9~$\mu$m results obtained with the Dwek \& Arendt method using the average of the best current direct determinations of the 2.2~$\mu$m CIB discussed below (22$\pm$6~\nwat) rather than the integrated galaxy light.  

Gorjian et al (2000) reduced the stellar foreground uncertainty directly by measuring all of the stars brighter than 9th magnitude at 2.2 and 3.5~$\mu$m in a dark 2$^\circ$ x 2$^\circ$ field near the north Galactic pole.  They calculated the contribution of fainter stars using the statistical model of Wainscoat et al (1992).  They also used a model for the zodiacal light contribution which differed from that of Kelsall et al (1998).  They argued that the Kelsall et al model left a high Galactic latitude residual at 25~$\mu$m which is dominated by IPD emission. This high 25~$\mu$m residual is unlikely to be Galactic cirrus because the brightness relative to the cirrus at 100~$\mu$m is much too high (as indicated by the ISM color analysis of Arendt et al 1998), and it is unlikely to be the CIB because it violates upper limits set by TeV $\gamma$-ray attenuation ($\S$~\ref{tev_lims}).   Gorjian et al therefore added a constraint to their IPD model that the residual at 25~$\mu$m after zodiacal light removal be zero at high Galactic latitude (the ``very strong no zodi principle'' of Wright 1997; Wright 2001a).  Their IPD model was otherwise similar to that of Kelsall et al, with parameters determined by fitting the apparent annual variation of the sky brightness.
Gorjian et al found significant positive residuals at 2.2 and 3.5~$\mu$m which they identified as probable detections of the CIB (Table~\ref{cib_tab}).  Had they used the IPD model of Kelsall et al, their residual emission would have been $\sim$40\% higher, clearly illustrating the lack of uniqueness in the IPD model.  With a field covering only $\sim$8 DIRBE beams, they did not demonstrate the isotropy of the tentative CIB signal.

Wright \& Reese (2000) used a distinctly different approach to determine the CIB at 2.2 and 3.5~$\mu$m.  At each wavelength, they compared the histogram of the pixel intensity distribution in the DIRBE  map with the histogram predicted from the star count model of Wainscoat et al (1992).  The IPD contribution had first been removed using the model of Gorjian et al. They found that the predicted histograms had the same shape as those observed, but had to be displaced by a constant intensity to agree with the observed histograms.  The shift was consistent in five fields at high Galactic and ecliptic latitudes.  They interpreted this shift as the CIB, and obtained average values  for the five fields consistent with the values found by Gorjian et al (Table~\ref{cib_tab}).  They noted that the histogram method is statistically more powerful for finding a real residual and less subject to systematic errors in the star count model than the subtraction approach used by Arendt et al (1998).  

The case for detection of the CIB at 2.2~$\mu$m was considerably strengthened by Wright (2001b).  He used data from the initial catalog from the Two Micron All Sky Survey (2MASS) (Cutri et al., 2000) to remove the contribution of Galactic stars brighter than 14th magnitude from the DIRBE maps at 1.25 and 2.2~$\mu$m in four dark regions in the north and south Galactic polar caps.
Each region was about 2$^\circ$ in radius.  Using the same IPD model as Gorjian et al, Wright obtained 2.2~$\mu$m residuals in his four fields consistent with  each other and with those of Gorjian et al and of Wright \& Reese (Table~\ref{cib_tab}).  Hence, there is a reasonable case for detection of the CIB at 2.2~$\mu$m, but its brightness has substantial uncertainty due to the uncertainty in the IPD contribution and isotropy has been demonstrated only over a very limited sky area.  The scatter in the residuals at 1.25~$\mu$m was too large to claim a detection.  Wright's analysis provides the strongest current upper limit on the CIB at 1.25~$\mu$m.  

\subsubsection{SUBMILLIMETER BACKGROUND BASED ON FIRAS DATA}
\label{firas_meas}
\begin{figure}[tbh]
\epsfxsize=4.75in	
\epsfbox{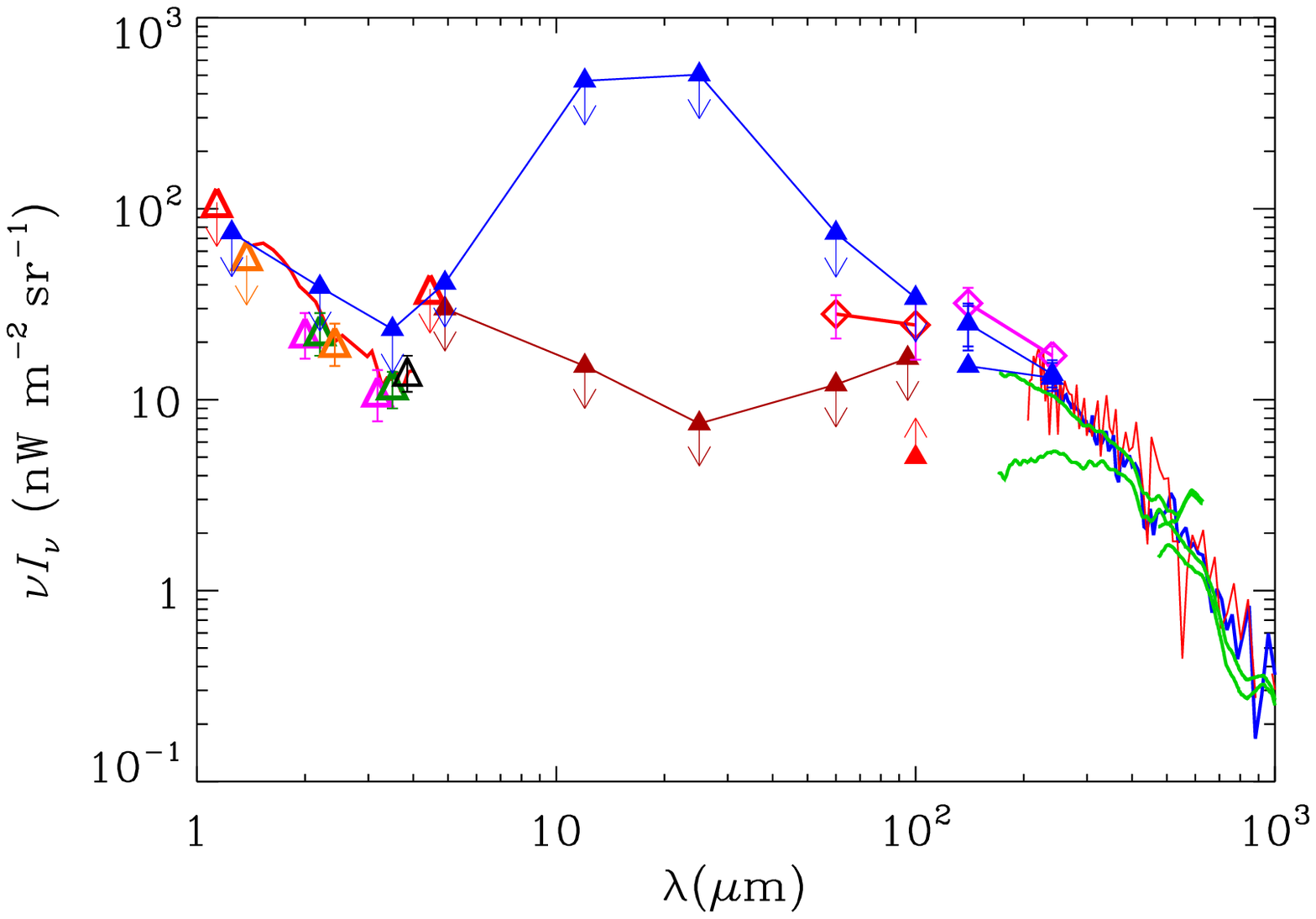}
\caption{}
\label{cib_plot}
\end{figure}

Puget et al (1996) reported the first tentative identification of the CIB using {\it COBE} data.  Analyzing the first released FIRAS data, they concluded that there was a residual uniform background from 200~$\mu$m to 2 mm in excess of contributions from the CMB, interplanetary dust, and interstellar dust.  They constructed a simple model for the IPD brightness based upon DIRBE 25 and 100~$\mu$m data and the assumption that the zodiacal emission is only a function of ecliptic latitude.  They extrapolated this model to wavelengths longer than 100~$\mu$m assuming $I_\nu \propto \nu^3$.  Since at these wavelengths the IPD contribution is small, an accurate IPD model is not required. They assumed that the ISM emission at high Galactic latitude and low H~I column density is traced by H~I, and that the submillimeter emission correlated with H~I can be represented by a single temperature medium with a $\nu^2$ emissivity law.  They also made a correction for infrared emission associated with ionized gas not correlated with the H~I, which significantly reduced their residual at the shortest wavelengths (200--400~$\mu$m).  Their residual maps had no significant gradients with Galactic latitude or longitude.  Figure~\ref{cib_plot} shows the spectrum obtained by Puget et al, with and without the ionized gas correction.  

Fixsen et al (1998) carried out a more extensive analysis of the FIRAS data, based upon the final photometric reduction of those data.  They subtracted the final FIRAS determination of the CMB (Fixsen et al 1996) and the IPD contribution obtained by extrapolation of the model of Kelsall et al (1998).  They used three different methods to look for an isotropic residual distinct from the ISM.  One method assumed that the ISM spectrum is the same in all directions, but the intensity is spatially varying.  The second method assumed that the neutral and ionized phases of the ISM are traced by a combination of maps of H~I 21-cm line emission and of C~II[158~$\mu$m] line emission.  The latter was mapped over the sky by FIRAS (Bennett et al 1994).  Each component, including a term proportional to the square of the H~I intensity, was allowed to have a distinct spectrum.  In the third method, they assumed that the submillimeter emission of the ISM is traced by a linear combination of the DIRBE ISM maps at 140 and 240~$\mu$m (Arendt et al 1998), each term again having its own spectrum.  Though each of these methods is subject to quite different systematic errors, the three methods yielded a consistent residual isotropic background, providing confidence that this is a robust determination of the submillimeter spectrum of the CIB.  An analytic representation of the mean residual from these methods is given in Table~\ref{cib_tab}, and the spectrum is plotted in Figure~\ref{cib_plot}.  The results are consistent with the DIRBE results at 140 and 240~$\mu$m of Hauser et al (1998).

Lagache et al (1999) extended the study of the far infrared emission of the Galaxy at high Galactic latitude using FIRAS  data, finding a component of the emission not correlated with H~I emission but which follows a cosecant($b$) law.  They attributed this emission to the warm ionized medium in the Galaxy, and subtracted the infrared emission associated with this component, as well as that correlated with H~I, from the mean FIRAS spectrum in low H~I column density regions to obtain the spectrum of the CIB longward of 200~$\mu$m.  The result is consistent with that of Fixsen et al (1998) (Table~\ref{cib_tab}, Figure~\ref{cib_plot}).  Lagache et al also applied a similar correction to estimate the CIB at 140 and 240~$\mu$m from DIRBE data in the Lockman Hole region, and concluded that the results were significantly lower than those of Hauser et al (1998).   However, as shown in Table~\ref{cib_tab}, the 240~$\mu$m CIB intensity is within 1$\sigma$ of the Hauser et al value, and the 140~$\mu$m value, which differs by less than 2$\sigma$, is no longer positive by 3$\sigma$ and might best be interpreted as providing an upper limit of 27~\nwat.  The similarity between the results of these analyses is not surprising since Hauser et al (1998) had chosen fields for their best determination of the 140 and 240~$\mu$m CIB that were known to have low infrared emission from the ionized gas phase (less than 5 and 2~\nwat ~at 140 and 240~$\mu$m respectively.  See $\S$~\ref{dirbe_meas} and Arendt et al 1998).

Lagache et al (2000) extended the study of infrared emission from 
the warm ionized medium using data from the 
Wisconsin H$\alpha$ Mapper (WHAM) sky survey of Reynolds et al (1998).  They analyzed diffuse emission regions covering about 2\% of the sky at high Galactic latitude.  Assuming a constant electron density so they could relate emission measure to H$^{+}$ column density, they found a marginally better correlation of infrared emission at FIRAS resolution with a linear combination of H~I and H$^{+}$ column density than with H~I alone.  Their results suggest that 20--30\% of the far-infrared emission at high Galactic latitudes is uncorrelated with H~I gas.  Their values for the CIB at submillimeter wavelengths are consistent with those of Fixsen et al (1998) and Lagache et al (1999).  Their revised 140 and 240~$\mu$m residuals based on DIRBE data are higher than those of Lagache et al (1999), and are essentially the same as those of Hauser et al (1998).  However, these residuals are positive by less than 3$\sigma$ due to the small sky area analyzed and the uncertainties in the WHAM data, and so most confidently provide upper limits (Table~\ref{cib_tab}).  They obtained a significantly positive residual at 100~$\mu$m similar to that of Hauser et al (1998), but did not demonstrate that this residual was isotropic.

\subsection{Background Measurements from \it IRTS}
\label{irts}
The most recent instrument to provide diffuse infrared background measurements is the Near Infrared Spectrometer (NIRS) (Noda et al 1994), which operated for 30 days in 1995 on the Japanese {\it Infrared Telescope in Space} (Murakami et al 1994; Murakami et al 1996; Noda et al 1996).  The {\it IRTS} was flown on the Space Flyer Unit spacecraft, which was launched into a 486 km altitude, 28.5$^\circ$ inclination orbit.  In order to maintain maximum possible offset angles of the field of view from the Sun and Earth, sky coverage during the mission was limited to 7\%.  The {\it IRTS} telescope was a heavily baffled,  on-axis, 15-cm Ritchey-Chr\'{e}tien system with four instruments sharing the focal plane.  The NIRS, a grating spectrometer, had an 8 arcmin square beam and covered the spectral range from 1.4 to 4.0~$\mu$m with 0.12~$\mu$m resolution.  There were no intermediate optical stops in the system to trap light diffracted or scattered from structures in the optical beam.  The NIRS contained a full beam cold shutter, which allowed frequent measurement of the instrumental zero point and responsivity.

Matsumoto et al (2000) recently provided a preliminary analysis of the NIRS data, reporting  detection of the CIB based upon analysis of the five days of data which were least disturbed by atmospheric, lunar, and nuclear radiation effects.  The sky area analyzed included Galactic latitudes from 40$^\circ$ to 58$^\circ$, and ecliptic latitudes from 12$^\circ$ to 71$^\circ$.  The NIRS could detect individual stars down to a fainter level  than the DIRBE ($\sim$10.5 mag at 2.24~$\mu$m).  Matsumoto et al used a later version of the statistical Galaxy model used by Arendt et al (1998) to calculate the contribution from stars below their detection limit (Cohen 1997).  They adopted the IPD model of Kelsall et al (1998) to calculate the zodiacal light, interpolating between the DIRBE wavelengths to the wavelengths of the NIRS measurements.  After subtraction of the IPD contribution, the residuals showed some remaining variation with ecliptic latitude, which they interpreted as evidence for a fairly isotropic background.  To obtain a quantitative value for the background at each wavelength, they correlated their star-subtracted brightness at each point with the IPD model brightness, and used the extrapolation to zero IPD contribution as a measurement of the CIB.  

The CIB intensities reported by Matsumoto et al near 2.2 and 3.5~$\mu$m are similar to the values found by Gorjian et al (2000) (Figure~\ref{cib_plot}).  At shorter wavelengths, the Matsumoto et al (2000) results continue to rise steeply to $\sim$65~\nwat~at 1.4~$\mu$m.  This is somewhat above the 95\% CL upper limit at 1.25~$\mu$m of Wright (2001b).  The NIRS result implies an integrated background energy over 
the 1.4--4.0~$\mu$m range of $\sim$~30~\nwat~(see \S~\ref{total_energy} for discussion of the integrated background energy).  

The preliminary report by Matsumoto et al (2000) does not provide detail regarding systematic uncertainties in their results.  For example, information such as the zero point uncertainty, limits on stray light, and uncertainty due to a shutter light leak (Noda et al 1996), would be valuable.  The largest uncertainty in the reported CIB values is attributed to the IPD model. Use of the Kelsall et al model introduces uncertainty associated with the relative photometric calibrations of the DIRBE and NIRS measurements, in addition to the intrinsic uncertainties in the model.  Though consistently calibrated reference stars were used for gain calibration of the two instruments, the relative surface brightness photometry can be discrepant due to systematic errors in determination of the beam shape and spectral response for each instrument, as well as random errors in the measurements of the response of each instrument to the standard reference stars.  Matsumoto et al (2000) noted that there is some evidence of calibration inconsistency in that the correlation between the star-subtracted brightness of the NIRS maps and the DIRBE IPD model brightness did not have unity slope at all wavelengths.  It will be very valuable to have a more detailed account of the NIRS analysis and its uncertainties.

\subsection{Fluctuation Analysis}
\label{fluc_anal}

If the CIB originates at least in part from discrete sources, then fluctuations in the number of sources in an observer's field of view will produce fluctuations in 
the measured background brightness.  Hence, measurement of fluctuations in the extragalactic background reveal information about the number and distribution of contributing sources.  Fluctuations can be characterized by the two 
dimensional autocorrelation function, C($\theta$), of the sky brightness or by the corresponding two-dimensional power spectrum.  

Fluctuation measurements do not directly yield the total background light.  However, at least two approaches have been used to constrain the infrared background light using fluctuation measurements.  For a given model of the source comoving luminosity density, three-dimensional correlation function, and spectral energy distribution, one can calculate both the autocorrelation function on angular scales of interest and the background.  Measurements of the fluctuations can then be used to constrain the background light in the context of such models.  An alternative approach is to use fluctuation measurements to constrain the behavior of source number counts below the detection threshold for individual sources, and then to integrate the counts to constrain the total light.  
 
Measurement of extragalactic background fluctuations in the infrared is somewhat easier than direct measurement of the CIB since it is not necessary to determine the zero point of the photometric scale.  However, measurement of CIB fluctuations otherwise has many of the same challenges as CIB measurement, since foreground sources and instrumental noise can be dominant contributors to the measured power spectrum or autocorrelation function of the sky brightness.  For potential CIB fluctuation detections, it is also necessary to demonstrate isotropy, that is, that the fluctuations are uniform over the sky.  Table~\ref{fluc_tab} summarizes results of studies of infrared background fluctuations, which we express as the rms value of measured fluctuations, $\delta\,(\nu$I$_\nu$), or as the source power spectrum, P$_S$, depending on what was reported by the investigator.  In either case, the reported values refer to the fluctuations in the extragalactic background.  If the investigator related the fluctuations to the CIB brightness, we show that as well.


An early search for fluctuations in the CIB at 2.2~$\mu$m was conducted by Boughn et al (1986), who used ground-based instruments and standard sky-chopping techniques to measure fluctuations.  They reported upper limits on the brightness fluctuations on scales from $10\,''$ to $30\,''$ and from $60\,''$ to $300\,''$, and 
used these limits to constrain models of young galaxy formation without attempting to state limits on the intensity of the CIB. 

Kashlinsky et al (1996a) studied the fluctuations in the DIRBE maps, and used this information to set constraints on the mean intensity of the CIB.  
They calculated the expected zero-lag correlation, C(0), from sources clustered like galaxies.  They found that C(0)$^{1/2} = \delta\,(\nu$I$_\nu)$ is of order 5\%--10\% of that part of the CIB arising from such sources, and that this result is rather insensitive to the details of their assumptions about the galaxy three-dimensional spatial correlation function and luminosity evolution.  The variance in the DIRBE maps at 1.25, 2.2, and 3.5~$\mu$m was found to be dominated by the contribution from Galactic stars.  Analyzing a few relatively dark fields at high Galactic and ecliptic latitude, they masked discrete sources and removed linear gradients.  They found quite different values of  C(0) in their different fields, and interpreted the lowest value as an 
upper limit on the CIB fluctuations.  In accord with their model analysis, they multiplied the limits on $C(0)^{1/2}$ by a factor of 10 and reported the results as implied upper limits on $\nu$I$_{\nu}$ arising from sources clustered like galaxies, which they assumed to be the dominant sources of the CIB (Table~\ref{fluc_tab}).

Kashlinsky et al (1996b) extended this type of analysis into the far infrared, including DIRBE wavelengths  of 100~$\mu$m and shorter.  They analyzed most of the sky (384 patches of $10^{\circ}$ x $10^{\circ}$~ each), first removing the IPD contribution with an early version of the model of Kelsall et al (1998).  After masking discrete sources and removing the smoothly varying background in each patch, they found that the fluctuations in the residual maps varied systematically over the sky, indicating that these fluctuations arose largely from residual foreground sources.  They reported the minimum value of the observed variance as an upper limit on the CIB fluctuations, finding the same upper limits  at 1.25, 2.2, and 3.5~$\mu$m as those of Kashlinsky et al (1996a).  They reported new limits on the fluctuations in the DIRBE bands from 4.9 to 100 $\mu$m, and again multiplied these limits by 10 to obtain upper limits on the background arising from sources clustered like galaxies.  This method of constraining the CIB is particularly appealing in this wavelength range, where the high brightness of the IPD makes direct measurement extremely difficult.

Kashlinsky \& Odenwald (2000) further analyzed the fluctuations in the DIRBE maps at 1.25--4.9~$\mu$m.   They found that the variation of C(0) could be fitted with a power law in csc($|b|$) plus a positive constant term, with    residuals showing no significant dependence on Galactic latitude or longitude in directions removed from strong Galactic signals.  Their analysis indicated that instrument noise, data reduction methods, and the Galaxy were not likely sources of the spatially constant component of the fluctuations.  They therefore suggested that these approximately isotropic fluctuations may arise from the CIB.   Kashlinsky \& Odenwald also re-analyzed the 12--100~$\mu$m data, obtaining upper limits on the residual fluctuations slightly lower than those reported by Kashlinsky et al (1996b).  Using their argument that the fluctuations are 5--10\% of the CIB, we have multiplied their fluctuation limits at these wavelengths by 15 to obtain the implied upper limits on the CIB shown in Table~\ref{fluc_tab}, results entirely consistent with those explicitly stated by Kashlinsky et al (1996b).

Matsumoto et al (2000) studied sky brightness fluctuations in the IRTS data.  After removing the discernible stars and large-scale features, they found a fluctuation spectrum which was dominated by read-out noise at wavelengths longer than 2.6~$\mu$m.  They did a Monte Carlo calculation of the expected fluctuations due to faint stars using the Galaxy model of Cohen (1997).  After removing the readout noise and calculated fluctuations from faint stars, they found positive excess fluctuations from 1.4 to 2.6~$\mu$m similar in magnitude to those reported by Kashlinsky and Odenwald (2000) (Table~\ref{fluc_tab}).  They also integrated their data from 1.4 to 2.1~$\mu$m, and calculated the spatial power spectrum along their observational strip, finding an indication of structure on scales of $1^{\circ}$--$2^{\circ}$.  This structure in the power spectrum agrees roughly with expected CIB fluctuations from galaxy clustering (Jimenez \& Kashlinsky 1999).

Somewhat contradictory evidence regarding CIB fluctuations in the near infrared was reported by Wright (2001b), who  presented upper limits on the CIB fluctuations at 1.25 and 2.2~$\mu$m.  Wright determined the standard deviation of the residuals after using 2MASS data to remove starlight from the DIRBE maps ($\S$~\ref{dirbe_meas}).  He interpreted the largest standard deviation in his four fields as an upper limit on CIB fluctuations, since there were potential non-cosmic contributions to the fluctuations which he ignored.  Wright's 1.25~$\mu$m upper limit is somewhat lower than the 92\% confidence interval for the detection of Kashlinsky \& Odenwald (2000) and well below the data of Matsumoto et al (2000) (Table~\ref{fluc_tab}).  His 2.2~$\mu$m limit is below the  detections reported by Matsumoto et al and Kashlinsky \& Odenwald, though within the 92\% confidence interval of Kashlinsky \& Odenwald.  

Burigana \& Popa (1998) searched for evidence of the CIB at submillimeter wavelengths  by looking for isotropically distributed fluctuations in the FIRAS maps.  They found consistent values of $C(0)^{1/2}$ in several fields at high Galactic latitude, and interpreted this as evidence for an extragalactic origin of the fluctuations (Table~\ref{fluc_tab}). 

Lagache \& Puget (2000) reported detection of CIB fluctuations based on analysis of deep 170~$\mu$m survey data in a $30'$ x $30'$ field (Marano~1 region) obtained with the ISOPHOT instrument on the {\it Infrared Space Observatory} ({\it ISO}) mission.  The primary foreground contribution to fluctuations at this wavelength is expected to be Galactic cirrus, which has been shown to have a steep power spectrum $P \propto k^{-3}$ in previous studies (Gautier et al 1992; Wright 1998; $k$ is the spatial frequency).  After removing sources brighter than 100 mJy from their map, the observed power spectrum varied as $k^{-3}$ for $k < 0.2$ arcmin$^{-1}$, but decreased more slowly in the range \mbox{0.25 arcmin$^{-1} < k <$ 0.6 arcmin$^{-1}$}.  They showed that the excess power above that expected for cirrus was much larger than instrumental noise, and attributed these fluctuations to unresolved extragalactic sources.  Since this observation was carried out in a single small field, they did not demonstrate isotropy of the fluctuation signal.  Their value is comparable to the upper limit found by Kashlinsky et al (1996b) and Kashlinsky and Odenwald (2000) at 100~$\mu$m.  It is about 5--10\% of the CIB brightness measurements at 140 and 240~$\mu$m (Table~\ref{cib_tab}), consistent with the model-based arguments of Kashlinsky and colleagues.

Matsuhara et al (2000) analyzed fluctuations in ISOPHOT 90 and 170~$\mu$m maps in two $44' \times 44'$ fields in the Lockman Hole.  As a result of this choice of field, fluctuations due to cirrus did not contribute appreciably to the measured power spectrum.  Their observed power spectra were flatter and brighter than expected for cirrus fluctuations.  The 90--170~$\mu$m color of the fluctuations was also warmer than that of cirrus.  They concluded that they had detected fluctuations from faint star-forming galaxies below their detection limits of 150 and 250 mJy at 90 and 170~$\mu$m respectively.  Adjusting their 170~$\mu$m result to the 100 mJy threshold used by Lagache and Puget (2000), they concluded that the 170~$\mu$m CIB fluctuations in the Lockman Hole field were consistent with those in the Marano~1 field.  While consistency in two fields does not constitute a strong isotropy test, it does provide further evidence that the reported fluctuations are indeed extragalactic in origin.
From these measurements, Matsuhara et al estimated limits on the behavior of source counts below their detection limits.  Using the source count constraints, they estimated the CIB contributions from sources brighter than 35 mJy and 60 mJy at 90 and 170~$\mu$m respectively.  The integrated 170~$\mu$m source light accounts for about 10-20\% of the measured CIB (Table~\ref{cib_tab}).

\subsection{CIB Limits from TeV $\gamma$-ray Observations}
\label{tev_lims}
Indirect evidence for the brightness of the CIB comes from observations of TeV $\gamma$-rays from blazars.  A beam of high energy photons can be strongly attenuated by 
e$^+$e$^-$ pair production through collisions with low energy photons (Nikishov 1962; Jelley 1966; Gould \& Schr\'eder 1966). The $\gamma + 
\gamma_b
\rightarrow e^+ + e^-$ interaction cross section of a $\gamma$-ray
photon of energy E$_{\gamma}$
with  isotropically distributed background photons $\gamma_b$ of
energy $\epsilon_b$ is rather 
sharply peaked at 
$\sim 1.5\times 10^{-25}$~cm$^2$ (Nikishov 1962). This
peak occurs at energies for 
which the product  E$_{\gamma}\epsilon_b
\approx 4\,(m_ec^2)^2 \approx$ 1~MeV$^2$, or $\lambda_b(\mu$m) $\approx$~1.24~E$_{\gamma}$(TeV) (Guy et al., 2000).  Hence, infrared photons
are particularly effective at attenuating $\gamma$-rays of energies
$\sim$1--100~TeV.

If the production of second generation $\gamma$-ray photons
appearing from the direction of
the source  can be neglected, then the observed $\gamma$-ray flux at
energy $E_{\gamma}$ from the source,
$J_{obs}(E_{\gamma})$, is simply related to the intrinsic source flux, $J_0(E_{\gamma})$, by
\begin{equation}
J_{obs}(E_{\gamma}) = J_0(E_{\gamma}) exp[-\tau_{\gamma \gamma}(E_{\gamma})],
\end{equation}
where $\tau_{\gamma \gamma}$ is the optical depth for $\gamma-\gamma$
interaction.
For a given E$_{\gamma}$, the optical depth is proportional to the
number density of
background photons. Very high energy $\gamma$-ray sources can therefore, in
principle, be used to probe the intensity of the CIB if they happen
to be relatively 
nearby (Stecker et al 1992).  Determination of the CIB from
$\gamma$-ray observations requires:  (1) knowledge of the intrinsic 
spectrum of the $\gamma$-ray  source; (2) knowledge of the spectral 
shape of the CIB; and (3) the assumption that the attenuation takes place in the intergalactic 
medium by the CIB instead of in the source itself or its immediate vicinity (B\"ottcher \& Dermer 1995;
Protheroe \& Biermann 1997; Bednarek \&  Protheroe 1999).

The probing of the CIB with very high energy $\gamma$-ray photons
has recently become practical with the development of imaging atmospheric \u Cerenkov telescopes, such
as the Whipple observatory, the High
Energy Gamma-Ray Array (HEGRA), and the \u Cerenkov Array at Themis
(CAT). The first extragalactic $\gamma$-ray source
detected at TeV energies was the blazar Markarian (Mrk)~421, located
at redshift $z~\simeq$~0.031 (Punch et al  1992). Two more sources, Mrk~501 and
1ES2344+514, at respective
redshifts $z~\simeq$~0.034 and $z~\simeq$~0.044, have been detected
since, but the latter source was only detected once, with marginal 
significance. Catanese \& Weekes (1999) have reviewed the 
$\gamma$-ray observatories and observations (see also McConnell \& 
Ryan 2000). Since all of the detected extragalactic TeV $\gamma$-ray 
sources are located at very low redshift, one can safely neglect evolutionary effects in
the CIB, which may be important in calculating the opacity to more distant sources (\S~\ref{gamma_opac}).

\begin{figure}
\epsfxsize=4.75in	
\epsfbox{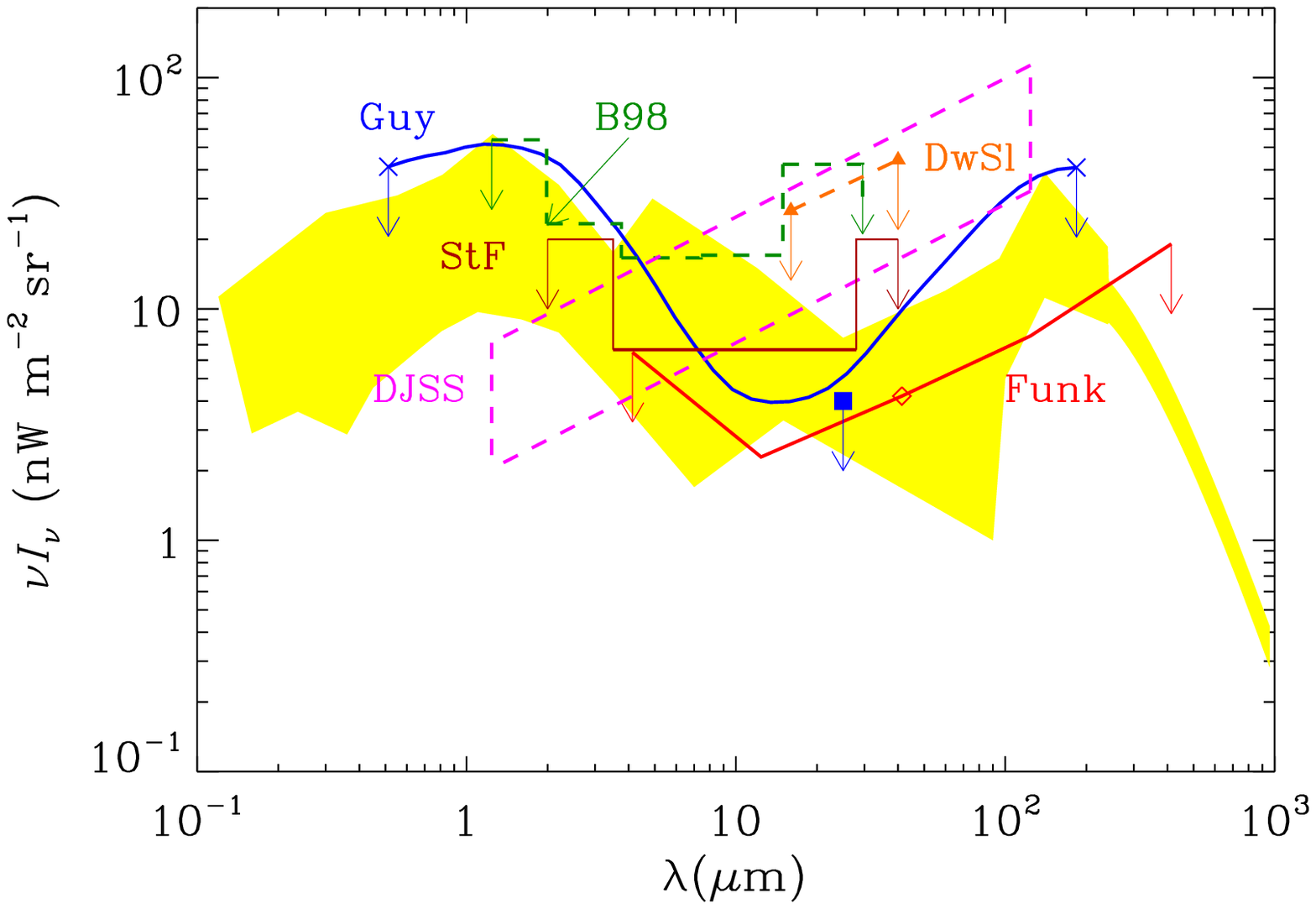}
\caption{}
\label{gammas_plot}
\end{figure}

In the first applications of  this technique, Stecker
\& De Jager (1993), De Jager et al (1994), and Dwek \& Slavin (1994), 
combined GeV observations of Mrk~421 obtained by the  Energetic Gamma Ray 
Experiment Telescope (EGRET) on board the {\it Compton Gamma--Ray Observatory}  ({\it CGRO}) (Lin et 
al  1992), with TeV observations obtained by the Whipple Observatory (Punch et al 1992), to 
infer the  intrinsic energy distribution of the source, and to search for evidence of a 
high energy cutoff in its  spectrum due to attenuation.  Background limits derived from their analyses are shown in Figure~\ref{gammas_plot}.
The problems of ascertaining reliable evidence for intergalactic 
attenuation were illustrated by Biller et al  (1995), who showed 
two equally acceptable fits to the spectrum of Mrk~421, the first requiring no 
attenuation, the second requiring significant CIB attenuation. Taking into account
the uncertainties in the extrapolation of the source spectrum from GeV to TeV energies and in the
spectral shape of the CIB, Biller et al  (1995) derived a very conservative upper limit of
$\sim$250\,$h$~\nwat~for the CIB at 10~$\mu$m.

Stanev \& Franceschini (1998) were the first to apply the attenuation 
analysis to Mrk~501, using high quality TeV $\gamma$-ray data obtained by the 
HEGRA experiment between March and October 1997 when the source was unusually bright. The
observed spectrum was fit by varying the  spectral index of the
$\gamma$-ray source spectrum and the  normalization of the CIB 
spectrum, which was 
taken to be the model spectrum of Franceschini et al  (1991). The 
best fit to the observations was consistent
with no absorption. A less model-dependent result was  obtained by
adopting the flattest allowable $\gamma$-ray source spectrum, and a flat $\nu I_{\nu}$
spectrum within the CIB wavelength bin that contributes most to the attenuation at a given $\gamma$-ray energy.  This approach yielded CIB upper limits, the most restrictive
being 7\,$h$ \nwat~in the \mbox{4--25~$\mu$m}
wavelength range. 

A similar approach was adopted by Funk et al  (1998), who fitted the Mrk~501
spectrum by an unattenuated
power law up to 10 TeV, suggesting that $\tau_{\gamma 
\gamma}$(10~TeV) $<$~1. By assuming a
spectral shape for the CIB based on the models of MacMinn \&
Primack (1996), they derived an upper limit of 4\,$h$~\nwat~at $\sim$40~$\mu$m, 
consistent with previous estimates. Biller et al  (1998)  derived new CIB upper
limits using minimal assumptions on the source
or the CIB spectra.  Approximating the CIB by a series of flat
$\nu I_{\nu}$ spectra in discrete wavelength intervals, they
calculated the $\gamma$-ray opacity
at various $\gamma$-ray energies. Constraining the overall index of the
differential photon spectrum of the source to be between 2.3 and 2.8, they derived an
upper limit of \mbox{16\,$h$~\nwat}~in the 4--15~$\mu$m interval.
Mannheim (1998) derived an upper limit on the
maximum allowed attenuation assuming a power law TeV $\gamma$-ray 
spectrum for Mrk~501, which implied a CIB limit of (3--5)\,$h$~\nwat~at 25~$\mu$m.

Recent simultaneous multiwavelength observations of Mrk~501 and 
Mrk~421 have  provided new constraints on the intrinsic source spectrum.  The observed double-humped 
nature of the spectra of Mrk~421 and 501, with peaks at X-ray  ($\sim$10~keV) and 
$\gamma$-ray ($\sim$1~TeV) energies, and the rapid, correlated variations of their X-ray and TeV 
spectra (Catanese \& Weekes 1999), are most  readily explained in the framework of the 
homogeneous synchrotron self-Compton (SSC)  model.  The SSC model and other 
competing models are described by Dermer et al (1997), Sikora (1997), and McConnell \& 
Ryan (2000). In the SSC model, the same electrons that produce the synchrotron radiation 
also upscatter the synchrotron photons  in the jet to TeV energies  by the inverse 
Compton process. The model produces a double peaked  spectral energy distribution  with a 
synchrotron (S) peak in the X-ray regime, and an inverse Compton (IC) peak in the
\mbox{$\sim$0.1--1~TeV} energy regime.
 
Guy et al  (2000) pointed out that the position and intensity of the IC peak in Mrk~501 is
affected by the $\gamma$-ray interactions with the optical and near-infrared photons of
the EBL. Adopting the spectral shape of the EBL predicted by Primack et al
(1999,  model LCDM; see \S~\ref{sa}), they calculated
the TeV opacity and reconstructed the source spectrum for various scaling factors for
the EBL intensity.  Making the general assumption that the flux of the source,
$\nu F_{\nu}$, should not rise between 4 and 17 TeV, and that its 
spectral slope at 600~GeV should be similar to that of the X-ray spectrum at 1~keV, 
they found that  the scaling factor must be less than about 2.5, setting CIB upper limits (1$\sigma$) of 
58\,$h$, $\sim$6\,$h$ and $\sim$35\,$h$~\nwat~at 1, 20, and 80~$\mu$m respectively.

Other studies have used the spectral observations, such as the
locations of the S and IC peaks, their variability, and the hardening
of the synchrotron spectrum, to constrain the parameter space of the SSC model.  
For example, Konopelko et al  (1999) argue that the intrinsic source
spectrum of Mrk~501 must be flat in $\nu F_{\nu}$ in order to comply
with the observational constraints, and therefore see the observed hardening of the spectrum
with energy as evidence for
intergalactic absorption. Adopting the Malkan \& Stecker (1998) model
for the EBL with the addition of a stellar emission component (the details of which have not been published), they derived an
essentially flat spectrum from 300~GeV to 20~TeV.  However, these results are at variance with
those derived by Guy et al (2000), presumably because of differences in the stellar emission models that 
produce the $\sim$1~TeV IC peak 
in the source spectrum. Sambruna et al  (2000) succeeded in modeling
the observed spectrum of Mrk~501 with the SSC model. Such a fit would imply no evidence for 
any intergalactic attenuation. However, the model over predicts the 10 TeV flux by a factor of 2.5, a discrepancy
regarded by them as only marginal evidence for intergalactic attenuation.

New upper limits on the CIB were derived by Dwek (2001, see Appendix) and Renault 
et al (2001), who explored the range of ``allowable" CIB spectra with 
only the minimal assumption that the intrinsic $\gamma$-ray spectrum of Mrk 501 be non-increasing at energies 
above $\sim\, 4$~TeV.  Dwek reported preliminary upper limits
of 7\,$h$ and 14\,$h$~\nwat~in the 6--30~$\mu$m 
interval and at 60~$\mu$m, respectively.  Renault et al found an 
upper limit of 7\,~$h$~\nwat~in the 5--15~$\mu$m interval.

Figure~\ref{gammas_plot} summarizes representative upper limits on the
CIB determined from TeV $\gamma$-ray observations. Dashed and
solid lines are limits determined from observations of Mrk~421 and
501 respectively.
Limits based on Mrk~501 are seen to be generally more restrictive. 
The lowest limits are 
in the 5--30~$\mu$m range since  the observed $\sim$1--15 
TeV~$\gamma$-rays interact most strongly with infrared photons in 
this wavelength range.  The shaded region in Figure~\ref{gammas_plot} is provided to allow comparison with other EBL measurements (see \S~\ref{ebl_sum}). 


The determinations of CIB intensities from
$\gamma$-ray  attenuation studies are somewhat model dependent, and 
are further uncertain due to the possibility that some attenuation 
may occur within the
$\gamma$-ray source.  However, even with the current 
limited data and uncertainties, the $\gamma$-ray observations are
providing valuable constraints on the CIB in the difficult 5--60~$\mu$m range.

\subsection{Integrated Light from Extragalactic Source Counts}
\label{source_cts}
Galaxy number counts provide important constraints on the background light. The cumulative brightness of galaxies is a strict lower limit on the background, and the number--magnitude relation provides important information on the nature and evolution of the sources contributing to the background. Conversely, the background light provides important constraints on galaxy number counts, providing an important measure of completeness, and limits on the possible existence of a truly diffuse background component.

A necessary condition for convergence of the integrated galactic light is that the logarithmic slope of the differential galaxy count per magnitude interval, $dlogN/dm$, should drop below a value of 0.4 at faint magnitudes. However, convergence does not ensure measurement of the total background, both because a significant fraction of the flux from galaxies can come from low surface brightness regions which are missed in standard galaxy aperture photometry, and because there may be truly diffuse sources of the background. The overlapping wings of resolved galaxies can create a relatively smooth unresolved background, which can only be detected by absolute surface photometry. We therefore consider integrated galaxy light as a lower limit to the background, even when the integrated light has converged. Background limits from galaxy counts are listed in Table~\ref{count_tab}.

Madau \& Pozzetti (2000) summarized deep galaxy counts in the {\it U, B, V, I, J, H} and {\it K} bandpasses compiled from the {\it Hubble Space Telescope} ({\it HST}) Northern and Southern Deep Fields (HDF-N and -S), supplemented with shallower ground-based observations in a variety of fields.  In all of these spectral bands, the logarithmic slope of the differential number counts has dropped below 0.4 at the depth of the HDF-S survey (m$_{AB}$ ranging from 25.5 at {\it K} to 30.5 at {\it V}), indicating that the light has largely converged.  [The AB magnitude system is defined by  AB=$-2.5\,\log F_{\nu}\, - 48.6$, where $F_{\nu}$ is in units of ergs\,cm$^{-2}$\,sec$^{-1}$\,Hz$^{-1}$.  The AB scale is chosen so that AB at 0.55 $\mu$m corresponds very nearly to the Johnson V magnitude (Oke 1974).]  The integrated light from these counts is presented in Table~\ref{count_tab}.

Gardner et al (2000) extended the deep galaxy counts in the HDF-N and HDF-S fields to shorter UV wavelengths using the Space Telescope Imaging Spectrograph (STIS).  At NUV (0.2365~$\mu$m) and FUV (0.1595~$\mu$m) wavelengths, these counts cover the brightness ranges of 23--29 and 23--30 AB magnitudes respectively.  The counts in both bands show a shallow or flat slope at the faint magnitudes sampled in these surveys.  To obtain estimates of the total galaxy light at these wavelengths, they combined their data with the counts of galaxies brighter than 20.75 AB magnitude at 0.2000~$\mu$m obtained by Milliard et al (1992) using the FOCA balloon-borne telescope.  


The IRAS Faint Source Survey reached sensitivity limits of $\sim$200~mJy at 12, 25, and 60~$\mu$m, and about 1~Jy at 100~$\mu$m. Hacking \& Houck (1987) analyzed the extensive IRAS survey and calibration data near the north ecliptic pole, reaching a depth of $\sim$~50 mJy at 60~$\mu$m.   Gregorich et al.\, (1995) analyzed a  subset of IRAS pointed observations, also reaching 60~$\mu$m sensitivity levels of 50--100 mJy. They found the source density to be about twice that reported by Hacking \& Houck.  However, Bertin et al.\, (1997) also analyzed the faint IRAS 60~$\mu$m counts to the 100~mJy level, choosing fields with very low cirrus, and found results inconsistent with those of Gregorich et al.\, and close to those of Hacking \& Houck.  They suggested that the counts in the areas studied by Gregorich et al.\, may have been elevated by false detections due to cirrus. We have integrated the infrared light from IRAS galaxy counts as summarized by Hacking \& Soifer (1991) at 25, 60 and 100~$\mu$m to the observational sensitivity limits  (Table~\ref{count_tab}).  The integrated light from the IRAS galaxies is far from converging, and is therefore a very conservative lower limit on the CIB.  

The ISOCAM and ISOPHOT instruments on the {\it ISO} satellite were used for many surveys (Lemke et al 2000). 
Altieri et al  (1999) carried out the deepest ISOCAM survey at 7 and 15~$\mu$m using the lensing cluster A2390,
reaching limiting fluxes down to $\sim$~30~$\mu$Jy. They combined the results of this
survey with counts obtained in 
the HDF (Aussel et al 1999; Taniguchi et al 1997; Elbaz et al 1999) to find 
the integrated intensity at 7 
and 15~$\mu$m (Table~\ref{count_tab}).  Clements et al (1999) conducted a deep 12~$\mu$m survey in four high
Galactic latitude fields with the ISOCAM, reaching a sensitivity of 0.5~mJy. 

Puget et al (1999) carried out an ISOPHOT survey at 175~$\mu$m, reaching a flux limit of $\sim$120~mJy in 
a 0.25 deg$^2$ area called the Marano~1 field.  To obtain the integrated galaxy light, Puget et al augmented their data with ISOPHOT 170~$\mu$m number counts from a survey of the Lockman Hole (Kawara et al 1998), and with counts
obtained by the ISOPHOT 170~$\mu$m Serendipity Survey (Stickel et al
2000).   Puget et al found the integrated light from sources brighter than 120~mJy to account for approximately 10\% of the detected CIB.   

Matsuhara et al (2000) reported the number counts and integrated light obtained with an ISOPHOT 90 and 170~$\mu$m survey of the Lockman Hole, with flux limits of 70 and 100 mJy
at the respective wavelengths. Juvela et al (2000) surveyed several fields with the ISOPHOT at wavelengths between 90 and 180~$\mu$m.  For sources brighter than 100 mJy detected at more than one wavelength, they found integrated surface brightnesses at 90, 150 and 180~$\mu$m similar to those  of Matsuhara et al and Puget et al (Table~\ref{count_tab}).

At submillimeter wavelengths, the rapid cosmological dimming of sources with increasing redshift is nearly compensated by the strongly negative K-correction for dusty sources.  That is, because the intrinsic source spectrum is brighter at wavelengths shorter than the observed submillimeter wavelength, one samples intrinsically brighter parts of the source spectrum for a large range of redshifts.  In principle, this makes dusty sources detectable to very high redshift (Blain \& Longair 1993).  Deep source counts at 850~$\mu$m have been obtained with the Submillimeter Common User Bolometer Array (SCUBA) on the James Clerk Maxwell Telescope. A population of about 20  distant galaxies has been detected at 850~$\mu$m down to a 3$\sigma$ limiting flux level of $\sim$2~mJy (Barger et al 1999a, and references therein).  Blain et al (1999b) extended the 850~$\mu$m counts down to a flux level of 0.5~mJy in surveys conducted through lensing clusters, accounting for nearly all of the measured CIB (Table~\ref{count_tab}). 

\subsection{UV-Optical Background Observations}
\label{uvo}
In order to relate the energy in the infrared background with that in the closely related UV-optical background ($\S$~\ref{total_energy}), we include a brief summary of UV-optical background measurements (Table~\ref{uvo_tab}).  Measurement of the extragalactic background at these wavelengths is also extremely difficult, and numerous conflicting results have been reported.  The reader is referred to  reviews by Bowyer (1991), Henry (1991), Leinert et al (1998) and Henry (1999) for more information.
  
The major obstacle to measurement of the extragalactic optical background, as in the infrared, is the bright foreground radiation.  From the ground, the dominant sources are atmospheric airglow and zodiacal light.  Additional foreground sources include starlight and diffuse Galactic light: optical light scattered by interstellar dust.  Most of the early investigations reported upper limits.  Dube et al (1979) ingeniously assessed the zodiacal light contribution by measuring the sky brightness in and outside of Fraunhofer lines in the solar spectrum.  Lillie (1972) used the Wisconsin Experiment Package (WEP) aboard the {\it Orbiting Astronomical Observatory 2} ({\it OAO-2}) to observe above the atmosphere.  {\it Pioneer 10} measurements reported by Toller (1983) were taken at 3 AU from the Sun, avoiding both of the strong diffuse foreground sources.  However, Toller's results were limited by the starlight contribution in the large field of view of the photometer (2.3$^\circ$ x 2.3$^\circ$).

Mattila (1976) introduced the novel technique of using a Galactic dark cloud as an opaque screen to chop on and off the extragalactic background, thus canceling most of the bright foreground contributions to the sky brightness.  However, diffuse Galactic light scattered from the cloud must still be removed.  Mattila initially reported detection of the optical background light with this technique.  However, his results are brighter than the upper limits set by other investigators, including Spinrad and Stone (1978) who used the same technique.  Mattila continued his measurements, and subsequently reported an upper limit close to the value of his earlier detection, which he attributed to Mattila \& Schnur (see Mattila 1990; also described by Leinert et al 1998).  Leinert et al (1998) discuss systematic corrections to both the Toller and Dube et al results (Table~\ref{uvo_tab}).


Bernstein (1999) has described a potentially significant new result in a preliminary report on detection of the diffuse EBL at 0.3, 0.55, and 0.8~$\mu$m.  These results were obtained by combining absolute sky brightness measurements made with the {\it HST} Wide Field/Planetary Camera 2 (WFPC2) with simultaneous spectrophotometry from the duPont telescope at Las Campanas Observatory (LCO) and the {\it HST} Faint Object Spectrograph.  The zodiacal light contribution was determined from the depth of the Fraunhofer lines in the sky background.  After excluding light from detected stars and a correction for diffuse Galactic light, these measurements yielded the EBL due to extragalactic sources fainter than V$_{{\rm AB}} = 23$, the brightest sources statistically well-represented in the small {\it HST} field-of-view (5 arcmin$^2$).   Bernstein added the integrated light from galaxy counts brighter than this limit to obtain the total optical background shown in Table~\ref{uvo_tab}.  This claimed detection of the EBL must be regarded as tentative since the uncertainties are substantial and the background has not been shown to be isotropic.

Bernstein also reported lower limits on the EBL determined by a simplified aperture photometry method applied to the ensemble of detected galaxies in the WFPC2 field in the range $23 < {\rm V_{AB}} < 28$.   Bernstein added the light from counts of galaxies with V$_{{\rm AB}} < 23$ to obtain the 2$\sigma$ lower limits to the total resolved galaxy light shown in Table~\ref{count_tab}.  Even though the Bernstein lower limits include only the light from detected faint galaxies, they are generally higher than the results of Madau \& Pozzetti (2000; Table~\ref{count_tab}), a result which Bernstein attributes to the systematic underestimation of individual galaxy brightnesses in traditional aperture photometry.  

The EBL has proven to be very difficult to determine at UV wavelengths.  Observations in the far ultraviolet have the advantage that the zodiacal light is not significant (Henry 1991: Figure 3).  However, terrestrial airglow and Galactic sources remain a challenge.  Martin et al (1991) used a spectrometer on the Shuttle UVX mission to measure the FUV (1400--1900~\AA) background by studying several directions with low H~I column density.  Since Galactic sources could contribute much of this background, they regarded their result as a firm upper limit to the EBL.  Brown et al (2000) measured the FUV (1450--1900~\AA) background using the STIS instrument on {\it HST}.  Excluding resolved sources and selecting data only from the night side of the orbit to eliminate airglow, they measured the diffuse background along a line of sight with low H~I and extinction.  Their result is somewhat brighter than that of Martin et al, and must also be regarded as an upper limit to the EBL (Table~\ref{uvo_tab}).  Murthy et al (1999) analyzed 17 years of data on the diffuse FUV (912--1100~\AA) background from the {\it Voyager} Ultraviolet Spectrometer (UVS), obtaining a very low 1$\sigma$ upper limit on the isotropic extragalactic continuum. However, Edelstein et al (2000) analyzed the same data, concluding that systematic uncertainties dictate a substantially higher upper limit to the EBL (Table~\ref{uvo_tab}).  Clearly, determination of the extragalactic UV background remains a difficult problem.

\subsection{Summary of Extragalactic Background Observations}
\label{ebl_sum}
As the preceding Sections show, there is now a great deal of direct and indirect observational evidence providing CIB upper limits, lower limits, and tentative or probable detections.  Figure~\ref{ebl_plot} summarizes this evidence, together with the data at UV and optical wavelengths to provide a more complete picture of the extragalactic background light.  References for the direct infrared background measurements and the limits from infrared fluctuation measurements are in Table~\ref{cib_tab} and Table~\ref{fluc_tab} respectively.  References for the UV-optical measurements are in Table~\ref{uvo_tab}.  Figure~\ref{ebl_plot} also shows the lower limits on the EBL found from the integrated light of resolved extragalactic sources.  References for these limits are in Table~\ref{count_tab}.  Missing from this figure for the sake of clarity are the CIB upper limits deduced from analysis of TeV $\gamma$-ray data, which are shown in Figure~\ref{gammas_plot}.  The dotted line in Figure~\ref{ebl_plot} indicates the nominal measurement where detections are reported, or a somewhat arbitrary intermediate value between upper and lower limits where there is no claimed detection.


Not surprisingly, the most convincing detections of the diffuse infrared background are in the spectral windows in the local foregrounds near 3.5~$\mu$m and longward of 100~$\mu$m.  At other wavelengths, the most certain and constraining results come from integrated galaxy light.   The limits inferred from the less direct measures, including fluctuations in the infrared sky brightness and attenuation of TeV $\gamma$-rays, are somewhat model dependent and therefore less certain.  However, they are particularly useful in the 5--100 $\mu$m range where direct measurements are so difficult.

It is notable that present detections of the EBL and the independently determined upper and lower limits are not in conflict with each other and, at some wavelengths, are not very far apart.  In particular, the deepest SCUBA counts at 850 $\mu$m (Blain et al 1999b) yield an integrated source light comparable to the CIB brightness determined from the FIRAS data.  Thus, the discovery of the submillimeter background and resolution of its sources seem to have occurred virtually simultaneously.  In the 7--180~$\mu$m range, the counts from the {\it IRAS} and {\it ISO} instruments are rising too steeply for the light to have converged, so the integrated light is necessarily below the actual CIB.  However, the lowest TeV $\gamma$-ray limits on the CIB are only modestly greater than the integrated light from the {\it ISO}  7 and 15~$\mu$m number counts, suggesting that the counts may be resolving a significant fraction of the CIB at these wavelengths. 

The situation in the optical and near-infrared (0.3--2.2~$\mu$m) is somewhat different. Although the galaxy counts have been extended to sufficiently faint levels that the integrated light has converged (Madau \& Pozzetti 2000), there remains an apparent gap between the integrated galaxy light and the reported EBL measurements.  However, there may be systematic underestimates by a factor of $\sim$2 in the integrated galaxy light due to photometry incompleteness at low surface brightness, as argued by Bernstein (1999).  Furthermore, the claimed EBL measurements in this range are less than 4$\sigma$ positive and are at most only slightly more than 2$\sigma$ above the integrated galaxy light.    In view of the potentially large systematic errors in both the integrated galaxy light and the extragalactic background determinations, one can not confidently conclude that there is a significant difference between these two measures.  Such a difference would imply the existence of unidentified discrete or diffuse sources contributing to the background light.  

The shaded region in Figure~\ref{ebl_plot} indicates conservative upper and lower limits on the spectral energy distribution of the EBL based on all available measurements and their uncertainties.  We use these limits in the discussions of implications (\S~\ref{impl}) and models (\S~\ref{models}).  

\section{COSMOLOGICAL AND ASTROPHYSICAL IMPLICATIONS}
\label{impl}
In this section, we discuss relatively direct implications of the EBL 
measurements.  Models for the origin of the EBL are discussed in 
\S~\ref{models}.
\subsection{Total EBL Energy}
\label{total_energy}
Important insights into the nature and evolution of the sources
contributing to the EBL can
be gained by examining its integrated energy and spectral energy distribution.
Examination of Figure~\ref{ebl_plot} suggests that the EBL spectrum 
has two maxima,
with peaks at both optical and far infrared wavelengths.
This supports a picture of a background consisting largely of 
redshifted primary radiations at
optical and near infrared wavelengths, and reradiated thermal dust
emission at far infrared wavelengths. The existence of the long 
wavelength peak provides
compellling evidence that the dominant luminosity sources are dusty.

We define $I \equiv \int I_{\nu}\, d\nu\,=\,\int \nu I_{\nu}\, 
d\ln{\nu}$ as the
frequency-integrated brightness of the cosmic background radiation.  In the following tabulation, we give first the value of the integral (in \nwat) of the nominal EBL shown by the dashed line in Figure~\ref{ebl_plot}.  We show in parentheses the range of values of the integral obtained using the upper and lower limits defined by the shaded area in Figure~\ref{ebl_plot}.
\begin{eqnarray*}
I_{{\rm stellar}} & = & 54 \qquad (19-100),\ \qquad \qquad  \lambda   = 
0.16-3.5~\mu{\rm m} \\
I_{{\rm dust}}    & = & 34 \qquad (11-58),\ \ \qquad \qquad  \lambda   = 
3.5-140~\mu{\rm m} \\
			& = & 15\pm2,\ \  \qquad \qquad \qquad  \qquad   \lambda  = 
140-1000~\mu{\rm m} \\
I_{{\rm EBL}}     & = & 100 \qquad (45-170), \qquad \qquad    \lambda  = 
0.16-1000~\mu{\rm m} \\
I_{{\rm CIB}}     & = & 76 \qquad (36-120),\ \qquad \qquad    \lambda  = 
1-1000~\mu{\rm m}.
\end{eqnarray*}
The quoted uncertainty in the 140--1000~$\mu$m integral is 1$\sigma$.  In presenting these 
integrated backgrounds, we have separated out the 3.5--140~$\mu$m 
range because there have been no direct detections of the dust emission in this spectral range. 

Thermal emission from dust dominates the EBL
spectrum at wavelengths longward of $\sim$3.5~$\mu$m, and constitutes 
about 48\% of the nominal EBL. However, values ranging from 20\% to 80\% are consistent with the measurement uncertainties defined by the shaded area.  The nominal percentage is 
larger than the $\sim$~30\% contribution of dust emission to the 
local (within $\sim 75\,h^{-1}\,$Mpc) luminosity density (Soifer \& Neugebauer 
1991), which suggests that the relative contribution 
of dust to the total energy output in the universe was higher in the 
past than at present.  However, given the substantial uncertainty in the EBL measurements, the possibility that this fraction may have been constant or even lower in the past cannot be ruled out.

To compare the integrated EBL energy density with
other cosmological energy budgets,  it is useful to define the dimensionless radiative energy density,
$\Omega_R$, as:
\begin{equation}
\Omega_{R}  =   {4\pi \over c}I /(\rho_c c^2),
\end{equation}
where $\rho_c$ is the critical mass density of the universe, with
\begin{displaymath}
\rho_c\,c^2  =  1.69 \times\ 10^{-8}\, h^2\, {\rm erg~cm^{-3}}.
\end{displaymath}
The energy density in the EBL is therefore  a small fraction of the 
critical energy density,
\begin{equation}
\Omega_{EBL} =
2.48 \times\, 10^{-6} h^{-2}\,I_{100},
\end{equation}
where $I_{100}\,\equiv \,I_{EBL}$/(100~\nwat\,) lies in the range 0.45--1.7 as indicated above.
The integrated CMB intensity, $I_{CMB}$, is 1000~\nwat (Mather et
al 1999), or \mbox{$\Omega_{CMB}$ = 2.48 $\times$\,10$^{-5}\ h^{-2}$.}
Hence, the total EBL energy is also a small fraction
of the radiant energy in the CMB: 
\begin{equation}
\Omega_{EBL} / \Omega_{CMB} = 0.10\,I_{100}.
\end{equation}

\subsection{Element Production and the Cosmic Star Formation Rate}
\label{lim_sfr}

The total EBL intensity is a measure of the bolometric energy output in the
universe. It can be expressed as the integral of the comoving 
luminosity density ${\cal L}(z)$ over redshift $z$ (Dwek et al 1998):

\begin{eqnarray}
    I  & = & \left({c\over 4\pi}\right)\ \int_0^{\infty} {\cal L}(z)
\left|{dt\over
dz}\right|{dz\over 1+z} \\ \nonumber
      & = & 9.63\times 10^{-8}\ h^{-1} \int_0^{\infty} \left[{{\cal L}(z)\over
({\rm L}_{\odot}\ {\rm Mpc}^{-3})}\right]
    H_0\left|{dt\over dz}\right|{dz\over 1+z} \qquad \nwat
\end{eqnarray}
where (Longair 1998: Chapter 7)
\begin{equation}
{\rm H}_0\left|dt/dz\right|  =
(1+z)^{-1}\left[(1+z)^2(1+\Omega_Mz)-z(2+z)\Omega_{\Lambda}\right]^{-1/2},
\end{equation}
$\Omega_M\equiv
\rho_M/\rho_c$ is the present mass density of the universe normalized
to the critical density, and
$\Omega_{\Lambda} \equiv \Lambda/3H^2_0$ is the dimensionless
cosmological constant.

If we make the assumption that most of the EBL energy is produced by fusion of hydrogen, 
then the comoving luminosity
density ${\cal L}$ can be directly related to the cosmic star formation and
element production rates.
Fusion of hydrogen into heavier elements liberates 0.7\% of the rest mass
energy.  The mass fraction of baryons in the form of hydrogen is denoted by $X$, and the decrement in this fraction due to nucleosynthesis in stars by $\Delta X$.  The luminosity density can therefore
be expressed in terms of $\dot\rho_{_{\Delta X}}$, the comoving mass
production rate of helium and
heavier elements:
\begin{eqnarray}
{\cal L}({\rm L}_{\odot}\ {\rm Mpc}^{-3}) & = & 0.007 
\dot\rho_{_{\Delta X}}c^2 \\ \nonumber
    & = & 1.0\times10^{11}\dot\rho_{_{\Delta X}}({\rm M}_{\odot}\ {\rm
yr}^{-1}\ {\rm Mpc}^{-3}).
\end{eqnarray}

The relation between ${\cal L}$ and the comoving star formation rate, 
$\dot\rho_*$,
depends on the stellar initial mass function (IMF), and is therefore 
not as robust as the relation between ${\cal L}$ and the metal 
production rate. It is given by the convolution
\begin{equation}
{\cal L}(t) = \int_0^t \dot\rho_*(\tau)
L_b(t-\tau)d\tau,
\end{equation}
where $L_b(t)$ is the bolometric luminosity per unit mass of a 
stellar population of age $t$.
For a constant star formation rate and a Salpeter IMF, $n(M) \propto M^{-2.35}$ for $0.1\ 
$M$_{\odot}\,<\,M\,<\,120\ $M$_{\odot}$, where $n(M)$ is the number 
of stars formed per unit mass $M$, the
relation is given by
\begin{equation}
{\cal L}({\rm L}_{\odot}\ {\rm Mpc}^{-3}) = 
(7-14)\times10^9\,\dot\rho_*\,({\rm M}_{\odot}\ {\rm yr}^{-1}\
{\rm Mpc}^{-3}),
\end{equation}
for populations of ages $10^8$ to $10^{10}$ years.  If we assume a constant star formation rate over cosmic history, $<\dot\rho_*>$,
then Equations 5 and 9 imply
\begin{equation}
<\dot\rho_*>({\rm M}_{\odot}\ {\rm yr}^{-1}\ {\rm Mpc}^{-3})   =
(0.18-0.37)\,h\ I_{100} \qquad {\rm
for}
\qquad
\Omega_M = 1,\ \Omega_{\Lambda} =0,
\end{equation}
similar to the result of Madau \& Pozzetti (2000).  In the local universe the comoving star
formation rate is only about 
\mbox{0.01~M$_{\odot}$\,yr$^{-1}$\,Mpc$^{-3}$} (Madau et
al 1998). Thus, the observed intensity of the EBL suggests an average
cosmic star formation rate that
is 10--40 times higher than the rate at the present epoch.

To get a simple estimate of the amount of metals produced in the
universe, we first assume that all
metals were produced in a burst of star formation at redshift $z_e$, and that all of the
stellar energy was instantaneously released. The 
expression for $I$
then simplifies to (Peebles
1993; Pagel 1997)
\begin{equation}
I_{EBL}  =  \left({c\over 4\pi}\right){ 0.007 \rho_{_{\Delta X}}c^2\over
1+z_e}, \\ \nonumber
\end{equation}
where $\rho_{_{\Delta X}}$ is the comoving mass density of consumed
hydrogen atoms. We can write $\Delta X
\equiv \rho_{_{\Delta X}}/\rho_b = (\Omega_{_{\Delta X}}/ \Omega_b)$,  where
$\rho_b$ and $\Omega_b$ are, respectively, the baryonic mass density
and its value as a fraction of the
critical mass density. The mass fraction of processed baryonic 
material can then be expressed in
terms of the EBL intensity as
\begin{equation}
    \Delta X = 3.54\times10^{-4}\left({1+z_e\over \Omega_b h^2}\right) I_{100}.
\end{equation}
The value of $\Omega_b$
derived from Big Bang nucleosynthesis calculations is
$\Omega_bh^2$ = 0.0192$\pm$0.0018 (Olive et al 2000). For an emission 
redshift of $z_e$ = 1, we get
\begin{equation}
\Delta X = (0.037\pm0.004)\,I_{100}.
\end{equation}
Hence, to get the observed EBL ($I_{100} = 0.45-1.7$) with an
instantaneous energy release at 
$z_e$~=~1, one needs to convert 2\%--6\% of the nucleosynthetic baryon 
mass  density into helium
and heavier elements.  
The range of $\Delta X$ values is comparable to the solar value, 
\begin{displaymath}
\Delta X(\odot) \equiv\ \Delta Y\ +\ Z_{\odot}\ \approx\ 0.06, 
\end{displaymath}
where
$\Delta Y \approx$ 0.04 is the difference  between the
solar ($Y = 0.28$) and the primordial ($Y = 0.24$) $^4$He mass fraction, and $Z_{\odot}$
= 0.02 is the solar metallicity.  Hence, the processing of the local ISM has been very similar to that of the average baryonic matter in the universe.

The instantaneous injection of energy is a
considerably  oversimplified description
of reality.  Even if all stars were formed instantaneously at some redshift
$z_e$, their energy output would be spread over some time interval (Equation 8)
that is at least as long as their main sequence lifetime (Madau \&
Pozzetti 2000). The EBL intensity will consequently fall off less 
steeply with increasing $z_e$ than the
$(1+z_e)^{-1}$ cosmic expansion factor, lowering the fraction of processed 
baryonic mass density
needed to produce the observed EBL intensity.

In reality, the cosmic star formation rate (CSFR) 
evolves with redshift.  The EBL intensity can be used to test 
the validity of proposed star 
formation histories.  For example, the CSFR determined from UV-optical 
observations of galaxies by Madau et al (1998) implies  a total extragalactic background in the 
0.1--1000~$\mu$m range of $\sim$30~\nwat\ (Dwek et al 1998).  This increases to 
about 47~\nwat~for the CSFR of Steidel et al (1999).  Even the larger of these values is only marginally consistent with the allowed range of integrated EBL.

In general, the rate of star formation in individual galaxies can be
inferred from a number of diagnostics based on the spectral energy
distribution and energy output from young massive stars (Kennicutt 1998). Combining these diagnostics into a
comprehensive picture  of the global star formation history
requires knowledge of the redshift of the galaxies
being studied, and verification 
that no significant energy release has been overlooked by
the use of a limited number of diagnostics. Hence, 
UV-optical observations of galaxies will a priori provide only a partial 
description of the true CSFR.  Indeed, the discovery of dust-enshrouded luminous infrared galaxies in SCUBA 850~$\mu$m surveys and follow-up studies of the redshift distribution of these sources show that a 
significant fraction of the star formation activity at high redshift  is taking place behind a 
veil of dust hidden from UV and optical observers (Hughes et al 1998; Barger et al 1998, 
1999b, 2000; Lilly et al 1999).   Additional evidence suggesting that a 
significant fraction of the star formation occurred at redshifts
$z>$1--2 was provided by the iron abundance determinations in galaxy 
clusters with the {\it Advanced Satellite for Cosmology and Astrophysics} ({\it ASCA}) 
(Mushotzky \& Lowenstein 1997; Renzini 1997), and by radio observations of galaxies 
(Haarsma et al 2000; \S~\ref{radio}). Revised estimates of the CSFR as a function of 
redshift that take into account some of the more recent developments are presented by Blain et al (1999c) and Haarsma et al (2000).

Since a significant fraction of the energy released by stars is 
ultimately emitted at far-infrared wavelengths, it is interesting to examine what 
constraints can be set  on
the CSFR from the CIB detections alone. Equation~14 (\S~\ref{models}) 
shows the relation between the EBL spectrum and  the spectral luminosity density, ${\cal 
L}_{\nu}(\nu, z)$, which,
integrated over frequency, is proportional to the CSFR. The formal 
determination of the 
CSFR from the EBL is an inversion problem.  
Gispert et al
(2000) examined the possibility of determining the CSFR solely from the CIB
measurements  longward of 140~$\mu$m. To simplify the problem, 
they assumed 
that the intrinsic spectral luminosity, $L_{\nu}(\nu)$, of the galaxies that 
dominate the CIB does not evolve with time, so that ${\cal 
L}_{\nu}(\nu, z)$ can be written as the product
${\cal L}_{\nu}(\nu, z) = L_{\nu}(\nu) \Phi(z)$, where $\Phi(z)$ is the
comoving number density of galaxies. Gispert et al used Monte Carlo
simulations to explore a range of possible solutions for $\Phi(z)$
for  various spectral shapes and intensities of $L_{\nu}(\nu)$. 
The method provided only weak constraints on the bolometric luminosity density ${\cal L}(z)$ at
$z<$~1 because of the lack of CIB data at wavelengths shorter than
140~$\mu$m.  Their CSFR in the redshift range $z~\approx$~1--4 was
found to be consistent with that inferred from galaxy observations 
(Steidel et al 1999; Haarsma et al 2000).

Gispert et al (2000) concluded that the far-infrared region of the
CIB strongly constrains the CSFR in the redshift range $z \approx$~1--4.   However, Dwek et al (1998) showed
that with different assumptions on the spectra of the CIB sources and the stellar initial
mass function, one can obtain the observed CIB spectrum at wavelengths above 240~$\mu$m for two
distinctly different star formation histories (see \S~\ref{fe} and
Figure~\ref{models_fe}). The fact that different star formation
histories provide equally good fits to the far-infrared CIB shows that only 
deep surveys that resolve the EBL into its individual sources can  
provide the definitive star formation history. 

Galaxy counts obtained by the {\it IRAS}, {\it ISO}, and SCUBA
observations have already provided useful constraints on the redshift 
evolution of the CSFR.
The {\it IRAS} 60~$\mu$m counts suggest strong luminosity evolution $\propto
(1+z)^{3.2\pm0.5}$ up to $z \approx$
0.2 (Bertin et al 1997). The {\it ISO} counts extended this
evolutionary trend to $z\approx$~1 (Puget et al 1999; Altieri et al 
1999; Elbaz et al 1999). The SCUBA observations of
dust enshrouded sources at $z\geq$ 2 suggest that the CSFR at these 
redshifts should be moderately declining as $(1+z)^{-1.1}$, or at most be constant. 
Otherwise, the integrated source light would
exceed the submillimeter background (Dwek et al 1998; Blain 
et al 1999a;  Takeuchi et al 2001).

\subsection{The Radio--Infrared Background Connection}
\label{radio}
{\it IRAS} data revealed a remarkable correlation between the radio and far-infrared fluxes of galaxies, stretching over four orders of
magnitude in infrared flux density (Helou 1991).
The correlation holds separately
for the thermal and synchrotron components of the radio emission as
well as with their sum (Price
\& Duric 1992). In hindsight, this correlation
should not be surprising because the two
fluxes arise from the formation of massive stars (Lisenfeld et al 1996). These stars do most of the heating of the dust that emits the far-infrared radiation.  They also form H~II regions, which dominate the radio thermal emission, and, after their explosive death, 
accelerate the high energy particles which produce the synchrotron emission.  The correlation
does not hold for active galactic nuclei (AGN)-dominated galaxies, in which the radio emission is not associated with massive star formation.

Haarsma \& Partridge (1998) used the radio-infrared correlation to
estimate the contribution of star-forming galaxies to the cosmic radio background (CRB).  Assuming that the sources of the CIB are predominantly star-forming galaxies ($\S$~\ref{agn}),  and that the local radio-infrared relation can be extended to all star-forming galaxies, they found that star-forming galaxies account for about 50\% of the CRB at 170~cm, exclusive of the CMB contribution. They also showed that the integrated flux from
discrete radio sources associated with AGN is about half the CRB at 75~cm, providing a consistent model for the origin of the CRB.  

An important implication is that the cosmic star
formation history can be constructed from surveys of faint radio sources, avoiding the need to correct for the effects of dust extinction. Haarsma et al (2000) used this technique to study the star formation rate in the range $z~=~0-1.5$, confirming the rapid rise from $z$ = 0 to 1, but finding rates considerably higher than extinction-corrected rates based on optical data. 

\subsection{Interaction with High Energy $\gamma$-Rays}
\label{gamma_opac}
As discussed in \S~\ref{tev_lims}, the CIB strongly attenuates $\gamma$-rays from 1--100~TeV.
Since the cross section for the $\gamma$-$\gamma$ reaction varies
rather strongly with energy near the peak, $\sim$50\% of the total cross section for 17~TeV
photons arises from interactions
with 40--80~$\mu$m background photons (Guy et al  2000). If $\nu I_{\nu}$ = 
28~\nwat\ at 60~$\mu$m, as claimed by Finkbeiner et al  (2000), then the mean free path for 
17 TeV photons would be $\sim$14~Mpc.  This is considerably smaller
than the distance to Mrk~501 ($\approx$ 160~Mpc for $h$ = 0.65), which would imply that the intrinsic 17 TeV flux from Mrk 501 is larger than the observed flux  by a factor of $\exp(\sim12)\approx$~10$^5$. This leads to an
excessive power output for that galaxy, a situation described by Protheroe \& Meyer (2000)
as ``an infrared background-TeV $\gamma$-ray crisis".   

Several possible solutions have been suggested to resolve this ``crisis". (1) Harwit 
et al (1999) suggested that the actual flux of  10--20~TeV photons could
be much lower than inferred from the observations. This 
could occur if lower energy photons were to arrive coherently, simulating the effect of single 
10--20~TeV photons with the sum of their energies. (2) Kifune (1999) noted that because the photon energy-momentum relation violates Lorentz invariance in quantum gravity scenarios 
(Amelino-Camelia et al 1998, and references therein), the energy threshold for electron pair production can be raised, thus reducing the opacity  of the universe to TeV 
photons by orders of magnitude. (3) The most obvious solution to the crisis
is, of course, to assume that most of the 60 and 100~$\mu$m radiation tentatively identified as the CIB by Finkbeiner et al (2000) actually arises from foreground sources, a possibility suggested by the authors.

The opacity of the universe to TeV $\gamma$-rays as a function of
$\gamma$-ray energy and source redshift has been calculated using conventional physics by several authors (Primack et al 1999; Salamon \& Stecker 1998; Biller et al  1998). Photons arriving
from distant $\gamma$-ray sources traverse different intergalactic radiation fields
at different redshifts. Primack et al  used semi-analytical models (\S~\ref{sa}) to calculate the
evolution of the EBL with redshift, whereas Salamon \& Stecker calculated these evolutionary
effects in the framework of the Fall et al (1996) cosmic chemical evolution model (\S~\ref{cce}). For 
local sources, \mbox{$z\ll$ 1}, the opacity can be calculated from the local EBL 
(Coppi \& Aharonian 1999; Protheroe \& Meyer 2000). Figure~\ref{tau_plot}  presents the
opacity of the local universe for TeV $\gamma$-rays implied by the EBL measurements summarized in
Figure~\ref{ebl_plot}.  Figure~\ref{tau_plot} suggests that the uncertainties in 
the EBL are sufficiently large that the detection of 10--20 TeV 
$\gamma$-rays does not yet constitute a crisis, and that 30 TeV 
sources could potentially be visible up to a distance of 
$\sim$100\,$h^{-1}$ Mpc.


\subsection{Limit on Intergalactic Dust}
Intergalactic dust, if sufficiently abundant, could give rise to a 
truly diffuse emission component of the CIB. The possible existence of 
intergalactic dust, and the
consequences for determining the deceleration parameter of the 
universe, were first 
considered by Margolis \& Schramm (1977).
Recent observations of Type Ia supernovae at high redshift (Riess et al 1998; Perlmutter et al 1999), which imply either an accelerating universe or an unrecognized cause for dimming of the supernova light, have renewed interest in intergalactic dust.  Aguirre \& Haiman (2000) calculated the contribution 
such dust would make to the CIB, assuming it were heated by the ambient
intergalactic radiation field and sufficiently abundant to account 
for the dimming of the distant supernovae. They found
that such dust would produce most of the observed CIB at
850~$\mu$m.  However, discrete sources detected by the SCUBA survey account for 
almost all of the CIB at this wavelength, leaving little room for any diffuse emission component, and implying insufficient intergalactic dust to account for the supernova dimming.

\section{MODELS FOR THE ORIGIN OF THE CIB}
\label{models}
The integrated background intensity provides only limited constraints on the 
history of energy releases in the universe. The total intensity alone does not allow us 
to address detailed issues, such as the nature and evolution of the CIB sources, 
the relative contributions of AGN and star-forming galaxies at various wavelengths, 
and the history of 
star and element formation.  These issues can be addressed, in part, 
by considering the spectral energy distribution of the background.

The spectral intensity $I_{\nu}(\nu_0)$ of the EBL at the observed frequency
$\nu_0$ is given  by the integral over its sources (Peebles 1993):
\begin{equation}
I_{\nu}(\nu_0) = \left ({c\over 4 \pi}\right)\ \int_0^{\infty}
{\cal L}_{\nu}(\nu,z)\left|{dt\over dz}\right|dz,
\end{equation}
\noindent
where ${\cal L}_{\nu}(\nu, z$) is the spectral luminosity density of all
luminous objects and radiating particles in a comoving volume element at
redshift $z$, $\nu~=~\nu_0(1+z)$ is the frequency in the rest frame of the luminous
objects, and $\left|dt/dz\right|$ was given in Equation~6 
(\S~\ref{lim_sfr}).

In a dust-free universe, the spectral luminosity density, ${\cal L}_{\nu}(\nu, z$), can in principle be simply 
derived from a knowledge of the spectrum of the emitting sources and
the cosmic history of their energy release. In a dusty universe, the
total intensity remains unchanged, but the energy is redistributed over the
entire spectrum.  
Predicting this spectrum poses a  significant challenge since the
frequency distribution of the reradiated emission depends on a large number of
factors (Dwek 2001). On a
microscopic level, the emitted spectrum depends on the wavelength
dependence of the absorption and
scattering properties of the dust, which in turn depend on the dust
composition and size
distribution.  The reradiated spectrum also depends on the dust
abundance and the relative spatial distribution of energy sources and absorbing dust. Finally, the cumulative spectrum from all sources depends on evolutionary factors, including the history of dust formation and processes which destroy the dust, modify it, or redistribute it relative to the radiant sources.

Cosmic expansion reduces the overall  weight of the contribution from 
sources with $z\,>\,2$ (Harwit 1999).  However, luminous 
infrared sources at high redshift can dominate the CIB at submillimeter wavelengths because of the negative K-correction.  Nonnuclear sources,  while not contributing
significantly to the total EBL, may still be major contributors at specific wavelengths.  Models of the EBL must therefore be able to identify the contribution of the distinct energy sources to all parts of the spectrum. 

In the following, we first  discuss the contribution of AGN and other 
nonnuclear sources to the EBL (\S~\ref{agn}).  We then describe cosmic evolution models that provide estimates of  the nuclear contribution to the EBL (\S~\ref{ebl_spec}).

\subsection{The Contribution of AGN and Other Nonnuclear Sources to
the CIB}
\label{agn}
The energy output from AGN represents the major
nonnuclear contribution to the radiative energy budget of the 
universe exclusive of the CMB.  The energy of an AGN is derived from the release of gravitational energy associated with the accretion of matter onto a central black hole (BH) located in the nucleus of a host galaxy.

Simple estimates suggest that AGN are not the dominant contributors to the integrated EBL intensity.  Accretion onto a black hole releases about 10\% of the rest
mass energy of the accreted matter, which is large compared to the 0.7\% conversion efficiency of nuclear fusion.  However, the total amount of matter in central black holes comprises only $\sim$ 0.6\% of that in  stellar objects (Magorrian et al 1998; Fabian
\& Iwasawa 1999). The total energy released from AGN should therefore be about
10\%--20\% of that released from nuclear processes. However, AGN could
still contribute a significant fraction of the CIB intensity in specific wavelength regions.

The energy output of AGN spans a wide spectrum, ranging from radio
frequencies to $\gamma$-rays (Sanders \& Mirabel 1996; Grupe et al  1998).
Many AGN exhibit a thermal infrared excess over the non-thermal
continuum, associated with emission from hot ($\sim$60--100 K) dust (Haas et al 1998). At 
far infrared wavelengths, such sources manifest themselves as luminous or ultraluminous infrared galaxies (ULIRGs), first
recognized in the {\it IRAS} data (Soifer et al  1986). The infrared 
emission can be either powered by an AGN or by starbursts, and the relative contributions of these sources to the infrared emission in most objects is not well understood. Phenomenological criteria for distinguishing the energy sources are summarized by Genzel \& Cesarsky (2000). However, the interpretation of the observations is difficult, since source geometry, dust extinction, and the competition of dust for UV photons can affect diagnostics such as line ratios and line-to-continuum ratios (Fischer 2000).  
For example, NGC~6420, a galaxy first believed on the basis of such 
diagnostics to be primarily powered by starbursts, may in fact be predominantly powered by an AGN (Vignati et al  1999). 

While the relative contributions of starbursts or AGN to the energy budget of individual infrared galaxies is still unclear, recent developments in our understanding of the origin of the cosmic X-ray background (CXB) may clarify the global contribution of AGN to the spectrum of the CIB.  Since the discovery of the CXB (Giacconi et al  1962), understanding its origin has been a challenge.  The strong limits set by the {\it COBE}/FIRAS instrument on any {\it 
y}--distortion in the spectrum of the CMB (Fixsen et al  1996) essentially ruled out 
diffuse thermal emission from a hot intergalactic gas as a significant source of the 
hard CXB. Type 1 Seyfert galaxies, an unobscured (N$_H < $ 10$^{22}$ cm$^{-2}$) 
subclass of AGN, could account for the soft ($\sim$0.2--1~keV) X-ray background (Hasinger et 
al  1998; Schmidt et al  1998), but the same sources failed to account for the harder 
component of the CXB ($\sim$1--40~keV), which is flatter. 

This inability to account for the hard CXB with any superposition of known discrete extragalactic sources was a problem referred to as the ``spectral paradox" (see reviews by Boldt 1987; Fabian \& Barcons 1992). A new population of objects seemed to be needed to account for the hard X-ray background. Setti \& Woltjer (1989) suggested that a
superposition of strongly absorbed sources with an imposed low energy
cutoff in the $\sim$10~keV range could explain the hard CXB.
They argued that the existence of such sources is a natural prediction of
the unified theory of AGN (Antonucci 1993). Under the AGN unification scheme, the X-ray spectra of Seyfert~2 galaxies, which are viewed through large H-column densities, will be characterized by a sharp low-energy cutoff between 0.5 and 10~keV. The CXB could arise from a combination of unobscured Seyfert~1 galaxies and Seyfert~2 galaxies with different 
degrees of obscuration (Madau et al 1994; Comastri et al 1995, and 
further references therein). A recent deep X-ray survey with {\it Chandra} confirmed 
that about 75\% of the hard CXB originates from obscured AGN, lending support to the CXB models constructed within the unified theory of AGN (Mushotzky et al 2000). 

An immediate consequence of these results is that some fraction of the energy absorbed by the 
obscuring material must be absorbed by dust and re-radiated in the infrared.  This 
provides a link between the sources of the hard CXB and the CIB.  Since the SCUBA surveys at 850 $\mu$m have resolved most of the CIB, this linkage can be observationally explored by identifying the AGN contribution to the SCUBA sources.   

Steps towards that goal were recently taken by Ivison et al (2000), Severgnini et al (2000), and Barger et al (2001).   Ivison et al, using optical and radio data, found evidence for AGN activity in 3 of 7 SCUBA sources, though it was not clear whether the AGN activity dominated the submillimeter emission of these sources.  Severgnini et 
al correlated a sample of 850~$\mu$m SCUBA sources and 2--10 keV 
X-ray sources from {\it Chandra} and {\it BeppoSAX} observations at limiting fluxes that resolve more than 75\% of the background in 
each of the two energy bands. They detected only one SCUBA source among the hard X-ray 
sources, limiting the AGN contribution to the submillimeter background to less than 
$\sim$7\%. A significantly larger AGN contribution to the CIB would require a close association of the SCUBA sources with the fainter hard X-ray sources, which contribute at most 25\% of the hard CXB.

Barger et al (2001) conducted deep optical, near-infrared, SCUBA, 
and radio surveys centered on 20 hard X-ray sources detected in a flux-limited deep 
{\it Chandra} survey.  One of the X-ray sources was detected by the SCUBA survey, again 
implying that only about 10\% of the 850~$\mu$m CIB arises from such sources. 

The CXB-submillimeter connection has been recently explored theoretically in the context of the unified AGN model for the origin of the CXB by Almaini et al~(1999) and Gunn \& Shanks~(2001).  The models are essentially backward evolution models (\S~\ref{be}), in which the local X-ray luminosity density of AGN is evolved backward in time assuming pure luminosity evolution. Model parameters are constrained so that the AGN produce the hard CXB.
With somewhat different assumptions regarding the relation between infrared and X-ray luminosity in individual galaxies, both models conclude that AGN contribute $\sim$~10--20\% of the measured 850~$\mu$m background.  Because of the rather simple assumptions about the template infrared SED used in these models, the models are not particularly informative about the AGN contribution at wavelengths shorter than $\sim$100~$\mu$m.

A more physically oriented approach was used by Granato et al (1997), who constructed detailed radiative transfer models for calculating the infrared spectral energy distribution
(SED)  from circumnuclear dusty tori powered by an accreting BH, for 
a distribution of viewing angles. Adopting parameters that force these objects
to produce the hard CXB, Granato et al found that AGN contribute only about 1\% to the CIB intensity at 850~$\mu$m, substantially less than the value found with the aid of the recent hard X-ray observations. The differences between their results and those of Almaini et al (1999) and Gunn \& Shanks (2001) arise primarily from the different infrared spectra adopted for these objects, and different prescriptions for the evolution of 
their X-ray luminosity function.

The available evidence therefore strongly suggests that AGN contribute at most only 10\%--20\% of the CIB.  This conclusion is supported by the global energetic argument based on the abundance of black holes, by the observational studies of the relation between AGN, hard X-ray, and submillimeter sources, and by theoretical models for the CIB contributions from the sources of the hard CXB.

Brown dwarfs are other gravitational energy sources that might contribute to the CIB.  These are objects  whose mass falls below the minimum mass ($\sim$0.08~M$_{\odot}$) required for stable hydrogen burning (Burrows \& Liebert 1993).  Karimabadi
\& Blitz (1984) assumed that these objects radiate as blackbodies and 
calculated their evolutionary tracks on the H-R diagram.  Assuming that all the dark matter required to close the universe is contained in these objects, they found the brown dwarf contribution to the CIB to be at most 3 \nwat~in the 10- to 100-$\mu$m wavelength range.  Because the assumed amount of mass in such objects was greatly exaggerated, cooling substellar objects clearly make a negligible contribution to the CIB. 

An additional nonnuclear contribution to the EBL might arise from the radiative decay
of primordial particles.  By appropriately selecting the particle number density, particle mass, and decay redshift one can produce a wide range of 
background intensities at UV to far-infrared wavelengths (Bond et al 1986; Wang \& Field 1989; Sciama 1998).  However, lacking physical evidence to substantiate particular choices, predictions of background contributions from such mechanisms remain highly conjectural.

In summary, it appears reasonable to assume that the energy in the CIB arises largely from nuclear processes.  In the following, we review models in which nuclear processes provide the energy for the background radiation.

\subsection{The EBL Spectrum Expected From Nucleosynthesis}
\label{ebl_spec}
Numerous models have been developed for calculating the evolution of 
the cosmic spectral luminosity density, ${\cal L}_{\nu}(\nu, z$), as a function
of redshift, and thus for calculating  the EBL. Most of these models 
were originally used to  predict galaxy number counts in deep photometric surveys in order to probe galaxy evolution. In two review papers, Lonsdale (1995, 1996) grouped the various models into two
categories: backward  evolution (BE), and forward evolution (FE) models. We distinguish two additional
categories: semianalytical (SA) and cosmic chemical evolution (CCE) models. The  models  differ in their
degree of complexity, physical realism, and ability to account for observations or to make predictions. 
We give a brief description of models in each category, with references to more detailed accounts.

\subsubsection{BACKWARD EVOLUTION MODELS}
\label{be}
Backward evolution models extrapolate the spectral properties of
local galaxies to higher redshifts using some parametric form for their
evolution. In their simplest form, commonly referred to as no evolution (NE) models, these models assume
that neither the SED nor the comoving number density of galaxies evolve with time.
The spectral luminosity density, ${\cal L}_{\nu}(\nu, z) = {\cal L}_{\nu}(\nu, 0)$, is therefore explicitly
independent of redshift. At a specific frequency, $\nu_0$, it is
simply given by the product $L_{\nu_0}\times
\phi(L_{\nu_0}) dL_{\nu_0}$, where $L_{\nu_0}$ is the galaxy
spectral luminosity determined from observations of local galaxies, and
$\phi(L_{\nu_0})$ is the local galaxy luminosity function (LF). 
Implicit in the above  expression is a sum over galaxy types. The EBL is then obtained by integrating the local luminosity density up to a maximum redshift, $z_{max}$, the epoch when
galaxies are assumed to have first formed. Fundamental differences
exist between determinations of ${\cal L}_{\nu}(\nu_0)$ at UV, optical, and near-infrared wavelengths, where the
emission is primarily composed of starlight, and at mid- to far-infrared wavelengths, where
the  emission is dominated by thermal emission from dust.

At optical and near-infrared wavelengths, the galaxy LF is described by the
functional form given by Schechter (1976), with parameters that can
vary with galaxy morphology and spectral type (Binggeli et al 1988; Heyl et al  1997; Marzke \&
Da Costa  1997; Marzke et al  1998).
Galactic SEDs are constructed by fitting population synthesis models
(such as the models of Bruzual \& Charlot 1993) to
observed UV-optical spectra or photometric data for each
galaxy type (Pence 1976; Coleman et al 1980; Yoshii \&
Takahara 1988; Kennicutt 1992; Kinney et al  1996; Schmitt et al  1997; Jansen et al  2000).

The extension of this approach to $\lambda \geq 10\ \mu$m is complicated by
the fact that at these wavelengths the galaxy SED is dominated by thermal
emission from dust. The LF of infrared galaxies
differs fundamentally from the Schechter LF, which drops below its infrared
counterpart at luminosities above $\sim$10$^{11}$~L$_{\odot}$
(Rieke \& Lebofsky 1986; Soifer et al  1987; Isobe \&
Feigelson 1992). Functional forms and parameters for
the infrared LF derived from the {\it IRAS} survey were
presented by Fang et al
(1998; 12~$\mu$m), Shupe et al (1998; 25~$\mu$m), Soifer et al (1986, 1987;
60~$\mu$m), Yahil et al
(1991; 60~$\mu$m), Lawrence et al (1986; 60~$\mu$m), Saunders et  al 
(1990; 60~$\mu$m),
Rowan-Robinson et al (1987; 25, 60, and 100~$\mu$m), and Rush et al 
(1993; 12 and 60~$\mu$m). Luminosity functions derived from the {\it 
ISO} surveys were presented by Xu (2000) at 15~$\mu$m, 
and by Serjeant et al (2001) at 90~$\mu$m.
Comparison between the various luminosity functions reveals large
differences, especially at faint
luminosities (Saunders et al 1990; Malkan \& Stecker 2001), which can
result from bias due to large overdensities (the local Virgo super-cluster) in the sample, or due
to different functional forms used to fit the data.

Using infrared observations of galaxies in the local
universe, one can construct a library of galactic spectra that is
consistent with the 
observational constraints. The spectra need to include the emission
from the unidentified infrared
bands (UIB) at 3.3, 6.2, 7.7, 8.6, 11.3, and 12.7~$\mu$m and the 9.7 and 18~$\mu$m silicate absorption bands 
ubiquitously seen in the Galactic interstellar medium and in  the spectra of external
galaxies (Allamandola et al 1985; Roche et al 1991;
Lutz et al 1996; Helou et al  2000; Genzel \&
Cesarsky 2000; Sturm et al  2000). They also need  to be consistent
with the observed trend of increasing $S(60\ \mu$m)/$S(100\ \mu$m) and decreasing $S(12\ \mu$m)/$S(25\ \mu$m)
flux ratios with increasing infrared luminosity (Soifer \& Neugebauer 1991). SEDs consistent with these
color trends can be constructed from a linear combination of several
distinct emission components.  These are (1) a cirrus component, representing the emission from
dust and UIB carriers, the
latter most commonly identified with polycyclic aromatic hydrocarbon (PAH)
molecules residing in the diffuse atomic phase of the ISM and heated
by the general interstellar
radiation field; (2) a cold dust component, representing the emission
from dust residing in
molecular clouds; and (3) an H~II or starburst emission component,
representing the emission from
dust residing in H~II regions and heated by the ionizing radiation
field. An additional AGN
component may be needed to represent the spectra of some of the most
luminous infrared galaxies. Using
this simple procedure with two or more emission components, one can
reproduce the fluxes and colors
of {\it IRAS} galaxies with luminosities ranging from normal ($L
\sim 10^{8.5}\ L_{\odot}$) to the most luminous ($L \sim 10^{13}\
L_{\odot}$) galaxies (Rowan--Robinson
\& Crawford 1989; Beichman \&  Helou 1991; Pearson \& Rowan--Robinson
1996; Guiderdoni et al  1998;
Dwek et al 1998; Malkan
\& Stecker 1998, 2001; Xu et al 1998; Rowan-Robinson 2001; Dale et al 2001; Helou 2001).

Number counts predicted by no-evolution models are
often used as benchmarks with which observations are compared.
In general, the predicted counts are much lower
than the observed number counts, requiring abandonment of the simple assumptions made in these
models. For example,  {\it ISO} deep surveys show strong evolution in galaxy number counts compared
with NE model predictions at 12
$\mu$m (Clements et al  1999); 15~$\mu$m (Oliver et al  1997; Elbaz
et al  1999; Altieri et al
1999; Xu 2000); 90~$\mu$m (Matsuhara et al  2000); and 175~$\mu$m
(Puget et al  1999; Matsuhara
et al  2000). Evolution can be introduced into BE models as pure
luminosity evolution, manifested as a global scaling of galaxy spectra as
a function of redshift, or as pure density evolution, manifested as a
change in the comoving number density of galaxies with redshift. The  evolution is usually
characterized by a $(1+z)^{\gamma}$ dependence of the evolving
quantity, with values of $\gamma$ that can vary with redshift. Models using functional forms to characterize the
luminosity or density evolution and models that combine both types of evolution are summarized by Lonsdale (1996).

It is difficult to establish a unique interpretation of 
differences between observed  and predicted number counts in shallow 
surveys because of the coupling between the effects of luminosity and density evolution.  Number
counts to sufficient depth can be affected by the presence of 
spectral features. The UIB features, for
example, can move in and out of the survey band  with redshift (Xu et al 1998), as is the case with
the ISOCAM observations of the HDF (Aussel et al  1999). 
This aspect has been used by Xu (2000) to discriminate between luminosity and density evolution effects in
modeling the 15~$\mu$m number counts. The ISOCAM data suggest
luminosity evolution with $\gamma$ = 4.5, strongly ruling out pure density evolution. The  
presence of very luminous infrared galaxies at $z\,>$ 2 also cannot be reproduced by 
pure density evolution models (Blain et al 1999a).

In the model of Xu, and luminosity evolution models in general, all
galaxies are assumed to evolve. A more reasonable scenario is one in which only a fraction of the
galaxies evolve or, alternatively,  a fraction of the non-evolving galaxy population undergoes a
short-lived starburst phase. Such a more realistic evolutionary 
scenario was presented by Tan et al (1999). In the Tan et al model, galaxies are
explicitly divided into two distinct components: a disk component, and a starburst  component
representing galaxies undergoing mergers or strong tidal interactions. Parameters of
the LF differ for the two 
components, and are evolved
backward in time following distinct, observationally and physically
motivated prescriptions.   For example, the density of the starburst component is assumed
to evolve in proportion to the collision rate between galaxies.

The model of Rowan-Robinson (2001) uses a parametric approach to 
characterize the CSFR.  
It includes four  spectral components (cirrus, M~82 and Arp~220 type 
starbursts, and AGN
dust torus) to characterize the CIB sources, and an evolving 
60~$\mu$m luminosity
function, to create the needed mix of sources as a function of 
redshift.  The parameters
are determined by fitting   the local {\it IRAS} color-luminosity 
relation, the 60, 175,
and 850~$\mu$m number counts and the 140--750~$\mu$m CIB spectrum. 
The  model predicts source counts at optical, near- and mid-infrared wavelengths.  It 
also predicts the EBL spectrum and the contributions of each of the four components from UV 
to submillimeter wavelengths.

Figure~\ref{models_be} compares the EBL predicted by
representative BE models with current EBL observations.  The Hacking \& Soifer (1991)
model assumes pure luminosity evolution with
$\gamma$ = 3, $z_{max}$ = 3, and $h$ = 0.75. The Beichman \& Helou 
(1991) model assumes pure
density evolution with
$\gamma$ = 2, $z_{max}$ = 3, and $h$ = 0.75. Malkan \& Stecker (1998) assumed
pure luminosity evolution up to a 
redshift $z_{flat}$, with no evolution thereafter up to $z_{max}$ = 
4, and $h$ = 0.50. Figure~\ref{models_be} presents their  model with $\gamma$ = 3 and $z_{flat}$ = 2, which 
gives a background close to the DIRBE limits. Blain \& Longair (1993) adopted $h$ = 0.50, and a single 
temperature dust spectrum (60~K, model BL60; or 30~K, model BL30) to characterize the SED of {\it IRAS} galaxies. They used the 60~$\mu$m luminosity function of Saunders et al (1990), and
assumed pure luminosity evolution with $\gamma$ = 3 up to redshift $z$ = 2 and a constant luminosity at
earlier times.  Rowan-Robinson (2001) examined the EBL intensity for 
various cosmological models with $h\,$=\,1.0.  Figure~\ref{models_be} shows the model 
for $\Omega_M=0.3$ and $\Omega_{\Lambda}=0.7$. 
Figure~\ref{models_be} also shows the stellar contribution to the EBL calculated by  Yoshii \& Takahara (1988) for   $z_{max}$ = 5 and $h$ = 0.50.
All of the BE models shown in Figure~\ref{models_be}, except for that of Tan et
al, scale simply with the Hubble constant as H$_0^{-1}$.

The models in Figure~\ref{models_be} produce roughly similar values for the far-infrared
background. The Blain \& Longair 30~K dust model, and to a lesser extent the Beichman \&
Helou model, clearly overproduce the CIB at far-infrared wavelengths. The models of Blain
\& Longair,   Hacking \& Soifer, Beichman \& Helou, and Malkan \& Stecker were
published before the {\it COBE} detections and, except for the latter model, did not
attempt to predict the mid-infrared background. All models that calculate the
mid-infrared spectrum of the CIB give very similar results. The models of Xu and 
Rowan-Robinson include a broad spectral
feature, representing the cumulative redshifted UIB emission
features.

The main advantage of BE models is that they are simple, and offer a
quick comparison of observations with predicted galaxy number-magnitude, number-redshift,
color-magnitude, and other relations.  Their main disadvantage is 
that they are not constrained by the physical processes, such as star and metal formation, or 
radiative transfer processes, that go on in the galaxies they represent.

\subsubsection{FORWARD EVOLUTION MODELS}
\label{fe}
Some of the shortcomings of BE models are corrected in forward
evolution models.  At the  heart of these models is a spectral evolution code that 
evolves stellar populations and calculates the stellar, gas, and
metallicity content and SED of a galaxy as a
function of time starting at the onset of star formation. These models were pioneered by Tinsley (1974), and
are now widely
used to model and date the SED of globular clusters and various galaxy types
(Leitherer et al  1996 and references therein; Fioc
\& Rocca-Volmerange 1997; Jimenez
\& Kashlinsky 1999). Model input parameters include a prescription for the star
formation rate, the stellar IMF, and the chemical evolution.
The models rely on a wide
range of computational and
observational data sets, such as stellar evolutionary tracks,
libraries of observed and
calculated stellar atmospheres, stellar nucleosynthesis yields, and
the observed luminosity
functions of galaxies. Models are then set in a
cosmological framework by
specifying the values of
$H_0$, $\Omega_M$, and $\Omega_{\Lambda}$. Assuming an initial
formation epoch, these parameters
are used to map the temporal evolution of galaxies into redshift
space, allowing direct
comparison of the model predictions with observations. Model parameters are
adjusted to match the galaxy number counts, spectral energy
distribution, colors, and metallicity as a function of redshift.
In this sense, FE models are
essentially fancy BE models that
allow for a  consistent
``backward" evolution of galaxy parameters with time.

To predict the CIB at wavelengths larger than a few microns with FE 
models, spectral and chemical
evolution models must be generalized to include the effects of dust 
on the scattering, absorption, and thermal reradiation of starlight. 
They must also allow for the
possible evolution of the dust abundance, composition, size 
distribution, and other characteristics that may affect their optical 
properties (Franceschini 2001;
Dwek 2001). The single most important factor that determines the emerging UV to
far-infrared spectrum of galaxies is their opacity, which depends on
the distribution
of the dust relative to the radiation sources. Models for
different geometries of emitters and absorbers have been constructed by many authors.
They range in complexity from simple analytical prescriptions for the evolution of
dust opacity (Guiderdoni \& Rocca-Volmerange 1987; Mazzei et al 1994; Franceschini et al
1994; Wang \& Heckman 1996; Guiderdoni et al 1998), to simple
radiative transfer models
consisting of a uniform distribution of emitters and absorbers in
various geometric
configurations (Disney et al 1989; Calzetti et al 1994; Dwek \& V\'arosi 1996;
Devriendt et al 1999), to more complex models consisting of a clumpy 
distribution of absorbers, and a mix of uniformly distributed and partially embedded 
sources (Witt \&
Gordon 1996; Silva et al 1998; V\'arosi
\& Dwek 1999; Devriendt et al 1999; Misselt et al 2001;
Efstathiou et al 2000).

To determine the cumulative spectral contribution of
galaxies to the EBL, models must be able to follow the evolution of the opacity
and the reradiated infrared spectrum with redshift. Most models simply
assume that the opacity is proportional to the metallicity of the gas
to some power, and adopt an extinction law to characterize its 
wavelength dependence. In general, models adopt a Galactic extinction 
law (Mazzei et al 1994), or an
LMC extinction law for high redshift and presumably low metallicity
galaxies (Calzetti \& Heckman 1999), or an interpolation between the two laws,
reflecting the evolution of galaxy metallicity (Guiderdoni et al 1998). Detailed models show that the dust
composition, and hence the dust extinction law, should depend on the 
evolution of the dust-forming sources such as carbon stars,
OH/IR stars, and
supernovae (Dwek 1998; Todini
\& Ferrara 2001). Evolutionary effects on dust composition are most
important in pristine starburst galaxies, and less so in spiral galaxies (Dwek et 
al 2000; Dwek 2001).

Figure~\ref{models_fe} presents the CIB predictions of selected FE models.  The solid triangles represent the model
of Franceschini et al  (1994).  
The lines marked {\it UVO}, {\it
ED}, and {\it RR} show the models of Dwek et al (1998) for three 
distinct cosmic star formation histories. The {\it UVO} spectrum was calculated for the UV-optically
determined CSFR of Madau et al (1998) and falls short of
providing the total EBL intensity and the detected 140 and
240 $\mu$m fluxes. The spectra marked {\it RR} and {\it ED} were 
calculated for a CSFR in which a significant fraction of the star formation was assumed to 
take place behind a
veil of dust. Both spectra are consistent with the FIRAS CIB measurements,
illustrating that the far-infrared CIB spectrum alone does not uniquely determine the CSFR. 
The curve marked {\it CH} shows the model prediction of 
Calzetti \& Heckman
(1999). Realizing that the CSFR cannot be  reliably deduced from UV 
observations of heavily obscured galaxies, they developed a  special method
for reconstructing the intrinsic star formation rate of galaxies as
a function of redshift. The procedure did not provide a unique solution. One class 
of solutions required a large attenuation correction in the UV-determined CSFR at 
high redshift, and
produced a background intensity consistent with the current limits on 
the CIB. Calzetti
\& Heckman (1999) characterized the reradiated infrared emission by a single 
20~K temperature
blackbody with a $\lambda^{-2}$ emissivity  law. Their model provides 
therefore no
estimate of the CIB at mid-infrared wavelengths.


Forward evolution models are successful in fitting the SEDs of
individual galaxies, galaxy number counts in
select bands, and, as shown in Figure~\ref{models_fe}, the general characteristics
of the EBL.  However, a serious disadvantage of FE models is
the assumption of monolithic star formation. In these models,
galaxies form at the same time and evolve quiescently.  The models do 
not provide for galaxy interactions,
stochastic changes in the star formation rate, or 
morphological evolution of galaxies. In particular,
these models fail to match the
850~$\mu$m galaxy  number counts without including a new population
of ultraluminous infrared galaxies.
Such galaxies must be introduced in an ad hoc fashion in these models.

\subsubsection{SEMI-ANALYTICAL MODELS}
\label{sa}
Some shortcomings of the FE models can be addressed in part by using
semi-analytical
(SA) models for structure formation to predict the
observable characteristics of
galaxies and the intensity and spectrum of the EBL.
SA models provide a very useful formalism to study the
development of galaxies and
clusters of galaxies in a hierarchical scenario for galaxy formation
(see Cole et al
2000 for a review and further references).

Semi-analytical models must consider numerous physical processes in order to
reproduce observable galaxy
properties. These include the cooling of the gas that falls into the
halos, a prescription for the
formation of stars, a feedback mechanism that modulates the star formation
efficiency, a stellar IMF,
and a star formation efficiency during merger events. In addition, SA
models require the
standard stellar spectral evolution and chemical evolution models
that are used in FE models. The
models contain numerous adjustable parameters to match the observed
properties of galaxies in the
local universe (White \&
Frenk 1991; Kauffmann et al 1993; Cole et al 1994;
Somerville \& Primack 1999).

The use of SA models does not solve any of the fundamental problems
associated with calculating the extinction and SED of
galaxies. These quantities depend on dust parameters, on the geometry 
of emitters and absorbers, and on the evolution of these properties, 
none of which are provided by the models.  The main advantage of SA 
models is that, in spite of the many
adjustable parameters, they provide a physical approach to the 
formation and evolution of galaxies.  In particular, SA models decompose the
overall cosmic star formation rate into two distinct components, a 
quiescent one
representing the formation of stars in galactic disks, and a stochastic one
representing the contribution from bursts of star formation during
major galaxy interaction or merging events.
Because such interactions are expected to be
more common at high redshift, SA models should naturally be able to
reproduce  the SCUBA source counts, while maintaining the fits to the 
optical and
near-infrared number counts. The SCUBA sources are presumably the high redshift
counterparts of local ULIRGs, most of which are mergers. In practice, 
most SA models need 
to introduce an additional starburst component, or push model parameters to
their limit, in order to fit the {\it ISO} 175 and SCUBA 850~$\mu$m 
number counts (Guiderdoni et al 1998; Devriendt \& Guiderdoni 2000). Blain et al 
(1999a) found that to fit these number counts, starburst and AGN activity generated by mergers must proceed more efficiently and rapidly as redshift increases.

Figure~\ref{models_sacce} shows several models that use the 
semi-analytical approach to
calculate the EBL spectrum. The models of Devriendt \& Guiderdoni 
(2000) and Primack et
al (1999) use a 3-component dust model (\S~\ref{be}) to characterize the dust
emission, although the EBL spectrum of the latter authors shows no 
evidence of the spectral
discontinuities expected from the presence of UIB features. The Blain 
et al (1999a) model 
in Figure~\ref{models_sacce} was calculated for a single-temperature dust spectrum 
of 35~K, which provided the best fit to the far-infrared number counts. It is, 
therefore, not expected
to provide a realistic EBL spectrum at mid-infrared wavelengths.

Semi-analytical models incoporate a wide variety of physical processes to predict observed galaxy properties. In spite of their successes, there remain some discrepancies between model predictions and 
observations. The origins of these discrepancies  are often difficult to trace because of the inherent complexity of the models. Approximations used in describing the physical processes, uncertainties in the input data, and fundamental shortcomings of the approach could all contribute to the discrepancies.

\subsubsection{COSMIC CHEMICAL EVOLUTION MODELS}
\label{cce}
Because the extragalactic background light is an integrated measure of cosmic activity, summed over time and over the wide variety of processes and systems that have populated the universe, it can only inform us about global characteristics of cosmic history. A modeling approach which deals with average properties of the universe rather than the many complex details involved would most naturally, and perhaps most informatively, relate to the background radiation.  Cosmic chemical evolution (CCE) models use just such an approach, relating in a self-consistent way the time history of a few globally averaged properties of the universe.  The great advantages of the CCE approach are its global nature and intrinsic simplicity.  CCE models provide a picture of the evolution of the mean density of stars, interstellar gas, metals and radiation averaged over the entire population of galaxies in a large comoving volume element.  Inputs to the models are tracers of stellar activity (emitted light) and tracers of the ISM in galaxies (absorbed light).  Cosmic chemical evolution equations, analogous to Galactic chemical evolution models (Tinsley 1980; Pagel 1997), are solved to guarantee consistency between the global rates of interstellar gas depletion, star formation, and chemical enrichment.  The models do not require detailed knowledge of the complex processes by which galaxies actually form, assemble, and evolve. In particular, CCE models do not need to address explicitly the merger history of galaxies, the central theme in SA models.  Because of the global approach, CCE models do not predict galaxy number counts.
An informative review of CCE models has recently been presented by Fall (2001).

The CCE models themselves have evolved as available input data have improved (Fall \& Pei 1993; Pei \& Fall 1995; Fall et al 1996; Pei et al 1999).  In the most recent version (Pei et al 1999), primary inputs were the mean rest-frame UV emissivity as a function of redshift, as determined in recent deep optical surveys, and the ISM content of galaxies, as traced by H~I column density determined from quasar absorption line studies through damped Ly$\alpha$ systems (absorbers with H~I column density $N_{\rm HI} >  10^{20}$\ cm$^{-2}$).  Spectral synthesis models were used to calculate the spectral energy density of starlight at each redshift, as done in FE models. The LMC extinction law was adopted to calculate the opacity of grains  and, hence, determine the fraction of absorbed starlight at each redshift.  
One important property not provided by the quasar absorption line observations is the distribution of dust temperatures along the line of sight. A power law distribution of grain temperatures, which fits the local infrared luminosity density, was adopted to 
approximate the infrared emission spectrum of real galaxies. The DIRBE and FIRAS CIB measurements from 140--1000~$\mu$m were used as an input to determine a dust clumping parameter in the model. This established the relative amount of background light at short wavelengths (optical-near infrared) and long wavelengths (far-infrared).  The model then yielded the detailed shape of the EBL spectrum.  The model determined a self-consistent solution for the extinction-corrected comoving rate of star formation as a function of redshift, which is consistent with that inferred from recent H$\alpha$, SCUBA, and {\it ISO} 15 $\mu$m observations.  The model successfully reproduced numerous observations not used as input, including (1) the comoving rest-frame optical (0.44~$\mu$m), and near-infrared (1.0 and 2.2~$\mu$m) spectral luminosity densities in the $z\,\approx\,0-2$ redshift 
interval; (2) the 12, 25, 60, and 100~$\mu$m luminosity densities in the local universe; 
and (3) the mean abundance of heavy elements in damped Ly$\alpha$ systems for $z\,\approx\,0.4-3.5$. 

Figure~\ref{models_sacce} compares the EBL spectrum predicted by the 
Pei et al model with current measurements. The model clearly yields a double-peaked spectral energy distribution, with nearly equal energy in the short- and long-wavelength peaks.  This is consistent with the far-infrared and UV-optical emissivity data used as input to the model, though it clearly falls short of the recently reported UV-optical and near-infrared measurements.  The very low levels predicted in the mid-infrared are still consistent with observations.

\subsubsection{MODEL SUMMARY}
We have described a broad range of models, developed with different objectives and levels of detail. The value of the different approaches cannot be judged solely by their ability to reproduce the EBL measurements, since many were not primarily designed for that purpose.  The astrophysical input data evolve rapidly, causing even models that use the same approach to differ simply because they were produced at different times.  Nevertheless, most models described above  yield roughly similar CIB spectra from $\sim$5--1000~$\mu$m (Figures~\ref{models_be}--\ref{models_sacce}). The fundamental reason 
for this similarity is that most models incorporate similar cosmic star formation histories.
BE models implicitly assume a rising CSFR up to $z\,\sim$~1--1.5,
with a nearly constant rate at earlier times. FE and SA models 
attempt to reproduce the same 
characteristic CSFR in order to fit number counts or comoving 
spectral luminosity densities at different redshifts. Galaxy SEDs, 
another important ingredient in 
modeling the CIB, are based on locally observed galaxy spectra. 
Simple models, even single temperature dust emission models, can 
reproduce the far-infrared and submillimeter background, but are 
less realistic in the mid-infrared and at shorter 
wavelengths.  More complex models differ in the mix of galaxy types 
used at a given redshift, and in their treatment of the 
mid-infrared emission.  

Large differences between models occur in the UV-optical spectral 
range.  In general, BE models do not include the physical processes that link the CIB and the UV-optical spectrum. 
Those BE models that do include the UV-optical range accomplish this by incorporating template spectra. Other models arrive at a double-peaked EBL spectrum because they explicitly include the absorption of starlight and reemission by dust.  

What is apparent in all of these modeling efforts is that background 
measurements provide important constraints, highlighting deficiencies or weaknesses in some, 
requiring tuning in others.  As the background measurements continue 
to improve, they will move us further along toward a consistent 
picture of the formation of structure, metals, and background light.

\section{SUMMARY AND FUTURE DIRECTIONS}
\label{sum_fut}
In this review we have shown that knowledge of the cosmic infrared background, its sources and its implications has advanced dramatically in the past few years.  We now have claimed detections of the background at near-infrared and far-infrared/submillimeter wavelengths based on data from the {\it COBE} and {\it IRTS} missions.  Useful upper limits at other wavelengths come from these direct measurements and indirectly from observations of TeV $\gamma$-rays, whereas lower limits from ground--based galaxy counts and counts from {\it ISO} and SCUBA surveys further limit the observational uncertainty in the CIB (\S~\ref{ebl_sum}).  Although observational knowledge of the CIB still has much room for improvement, the detections and limits presented in this review already have
many important implications.

(1) The total intensity of the extragalactic background light from UV to millimeter wavelengths lies in the range 45--170~\nwat.  The energy in the 0.16 to 3.5 $\mu$m range is still quite uncertain, with present limits of 19--100~\nwat.   The energy in the 3.5 to 140~$\mu$m range,  11--58~\nwat, remains
similarly uncertain because of the dominance of the interplanetary dust emission. 
The most certain background determination is in the 140 to 1000~$\mu$m range, which contains 15$\pm$2~\nwat.  The nominal value for the total cosmic infrared background intensity (1--1000 $\mu$m) is 76~\nwat.  The nominal value for the total extragalactic background (0.16--1000 $\mu$m) is 100~\nwat.

(2) Using nominal EBL values, the energy in the background appears 
about equally divided between direct starlight (52\%) and starlight that has
been absorbed and reemitted by dust into the 3.5--1000~$\mu$m
wavelength regime (48\%).   However, the uncertainties in the present measurements allow the fraction from dust emission to range from 20\%--80\%.  

(3) The dominant energy source for the EBL from 0.16 to 1000 $\mu$m is apparently 
thermonuclear fusion reactions that convert hydrogen into heavier elements, whereas AGN contribute about 10\%--20\% of the total background.  If all of the nucleosynthetic energy release occurred at a redshift $z$=1, then a  fraction between 0.02 and 0.06 of the total baryonic matter was converted into elements heavier than hydrogen.

(4) The nominal fraction of the EBL intensity consisting of thermal emission
from dust (48\%) is higher than the $\sim$ 30\% found in the local 
universe, suggesting that dust-enshrouded objects made a larger contribution to the luminosity density in the past than at the present epoch.

(5) SCUBA observations at 850~$\mu$m identified such a population of luminous, dusty objects at high redshift.  The light from these objects is comparable to the 
CIB intensity at 850 $\mu$m, indicating that most or all of the background at this wavelength has been resolved.

(6) {\it ISO} source surveys from 7 to 170 $\mu$m reveal steeply rising counts of dusty sources at these wavelengths.  The {\it ISO} surveys near 170 $\mu$m have resolved $\sim$10\% of the measured CIB. 

(7) Because the EBL is an integral quantity, the measured EBL spectrum does not imply a unique star formation history.  Definitive determination of the CSFR history
will require direct observations of the sources that make up the EBL.  However, the integrated EBL implies that the average star formation rate over cosmic history is about an order of magnitude larger than the current rate.

(8) If the correlation between radio and far-infrared emission found in nearby galaxies applies to the star-forming galaxies that dominate the CIB, then these galaxies contribute about half of the cosmic radio background.  

(9) The CIB provides a significant source of opacity for ultrahigh
energy $\gamma$-rays due to electron-positron pair production.  
Blazars within distances of $\sim$100\,$h^{-1}$~Mpc
have therefore been used to probe the CIB, constraining the CIB 
intensity in the mid-infrared ($\sim$5--60~$\mu$m) spectral range. 

(10) We have discussed four distinct approaches that have been used to model the evolution of cosmic sources with varying degrees of physical realism and astrophysical detail.  Although most of these models were not primarily constructed to predict the CIB, they have been tuned to be generally consistent with the CIB intensity and spectral shape.  This tuning has provided insight into the history of star formation, metal production, and dust formation and distribution. 

We can anticipate that study of the CIB and its implications will continue to be vigorous for some time to come.  While there are no new space instruments under development for the specific purpose of diffuse infrared background measurements, one can expect there to be further improvements in limits or detections of the extragalactic infrared background based upon the sky brightness measurements already in hand and better determinations of the foregrounds.  Completion of the 2MASS survey at J, H and K bands will permit accurate removal of the stellar foreground over the substantial sky areas needed for convincing demonstration of isotropy.  The {\it Space Infrared Telescope Facility (SIRTF)} has the potential of providing direct measurements of foreground sources at 3.5~$\mu$m, though clearly over limited sky areas (Werner et al 2001).  More extensive maps of the distribution of H$^{+}$ will be available as the WHAM survey progresses, facilitating CIB discrimination particularly in the far infrared.  

The largest obstacle to direct extragalactic background light measurements over much of the infrared spectrum and at UV-optical wavelengths remains the bright foreground due to scattering and emission from interplanetary dust.  This major problem could be substantially eliminated by making measurements with instruments located several AU  from the Sun in the ecliptic plane, or in a location out of the ecliptic plane.  We are not aware of any current plans to accomplish such measurements.

Indirect constraints on the CIB from measurements of TeV $\gamma$-rays will continue to improve as more sources over a greater range of distances are observed.  Next generation TeV $\gamma$-ray telescopes, such as the Very Energetic Radiation Imaging Telescope Array System (VERITAS), will be able to detect fainter sources at higher
redshift,  providing needed confirmation of the intergalactic
absorption signature in the source spectrum (Catanese \& Weekes 1999).  Space missions such as the {\it Gamma-ray Large Area Space Telescope} ({\it GLAST}) will help to clarify the intrinsic spectrum of the TeV $\gamma$-ray sources from 20--200~GeV.  

There promises to be continued rapid advance in measurements of discrete extragalactic sources contributing to the background.  Such measurements will provide both increasingly robust lower limits to the CIB, and essential information on the distance and character of these sources.  Analyses of {\it ISO} data in the thermal and far-infrared are still in progress.  Source counts will be carried to much fainter levels by missions such as  {\it SIRTF} (Werner et al 2001), the {\it Far Infrared Space Telescope} ({\it FIRST}; Pilbratt 2001), and the {\it Infrared Imaging Surveyor} ({\it IRIS}; Shibai 2001).  The Atacama Large Millimeter Array (ALMA) will provide the sensitivity and angular resolution at submillimeter and millimeter wavelengths needed to clarify the nature of the sources being revealed in SCUBA observations.   Deep, wide-field near infrared imagery from the {\it Next Generation Space Telescope} ({\it NGST}; Stockman \& Mather 2001), combined with spectroscopy from {\it NGST} and large ground-based telescopes will dramatically advance our understanding of the nature and distribution of sources at high redshift and their contribution to the infrared background.  Within the next decade or so it should become clear whether the CIB arises entirely from discrete sources.

As observational knowledge of the cosmic histories of star formation and metal formation, and of the nature of the systems (AGN, starburst galaxies) making substantial contributions to the CIB improves, uncertainties in the models of background light generation associated with these processes will be reduced.  This offers the prospect of a consistent, comprehensive history of the growth of cosmic structure and accompanying energy releases. 

\section{ACKNOWLEDGMENTS}
During the preparation of this manuscript we have greatly benefited
from helpful and enlightening discussions with Rebecca Bernstein, Andrew Blain, Doug
Finkbeiner, Dale Fixsen, Alex Kashlinsky, Richard Mushotzky, Sten
Odenwald, Ray Protheroe, and Ned Wright. We thank Felix
Aharonian, Rick Arendt, Chuck Dermer, Mike Fall, Jim Felten, Michel
Fioc, Alex Konopelko, and Bob Silverberg, for their helpful comments on sections of the manuscript. We thank Prof.
T. Matsumoto for permission to reproduce the IRTS background spectrum.
We thank Andrew Blain, James Bullock, 
Daniela Calzetti, Julien Devriendt, Alberto Franceschini, Guilain
Lagache, Matt Malkan, Jonathan Tan, and Cong Xu for communicating
their model results in digital form. Finally, we thank the scientific editor, Allan Sandage, for his careful review and valuable comments.
Preparation of this review was partially supported by NASA grant NAG5--3899 and NASA contract NAS 5-26555 to the Association of Universities for Research in Astronomy, Inc. (MGH), and by NASA's Astrophysical Theory Program (NRA 99-OSS-01) (ED).

\newpage

\section{FIGURE LEGENDS}

\begin{figure}
\epsfxsize=4.75in	
\epsfbox{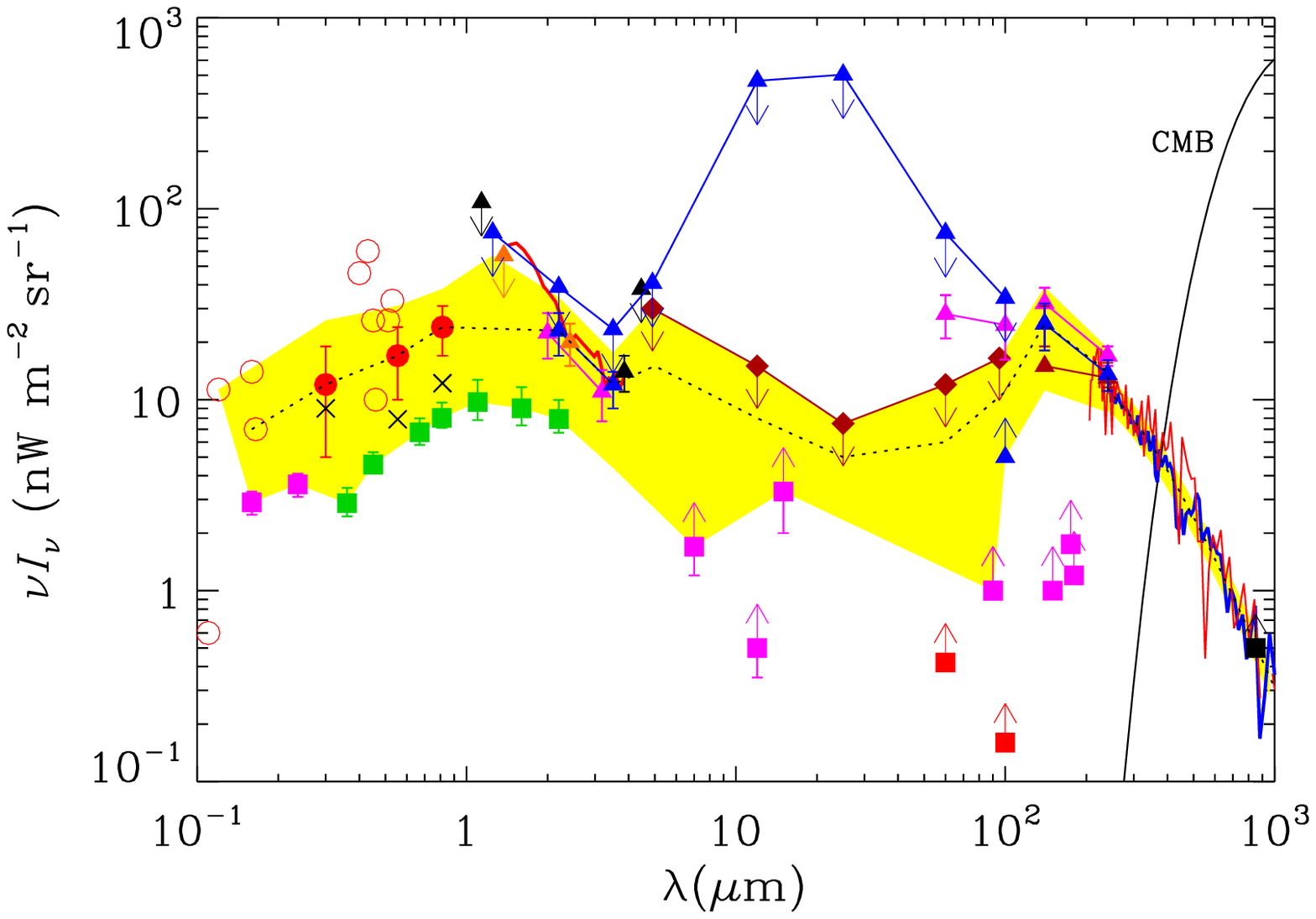}
\caption{}
\label{ebl_plot}
\end{figure}

\begin{figure}
\epsfxsize=4.75in	
\epsfbox{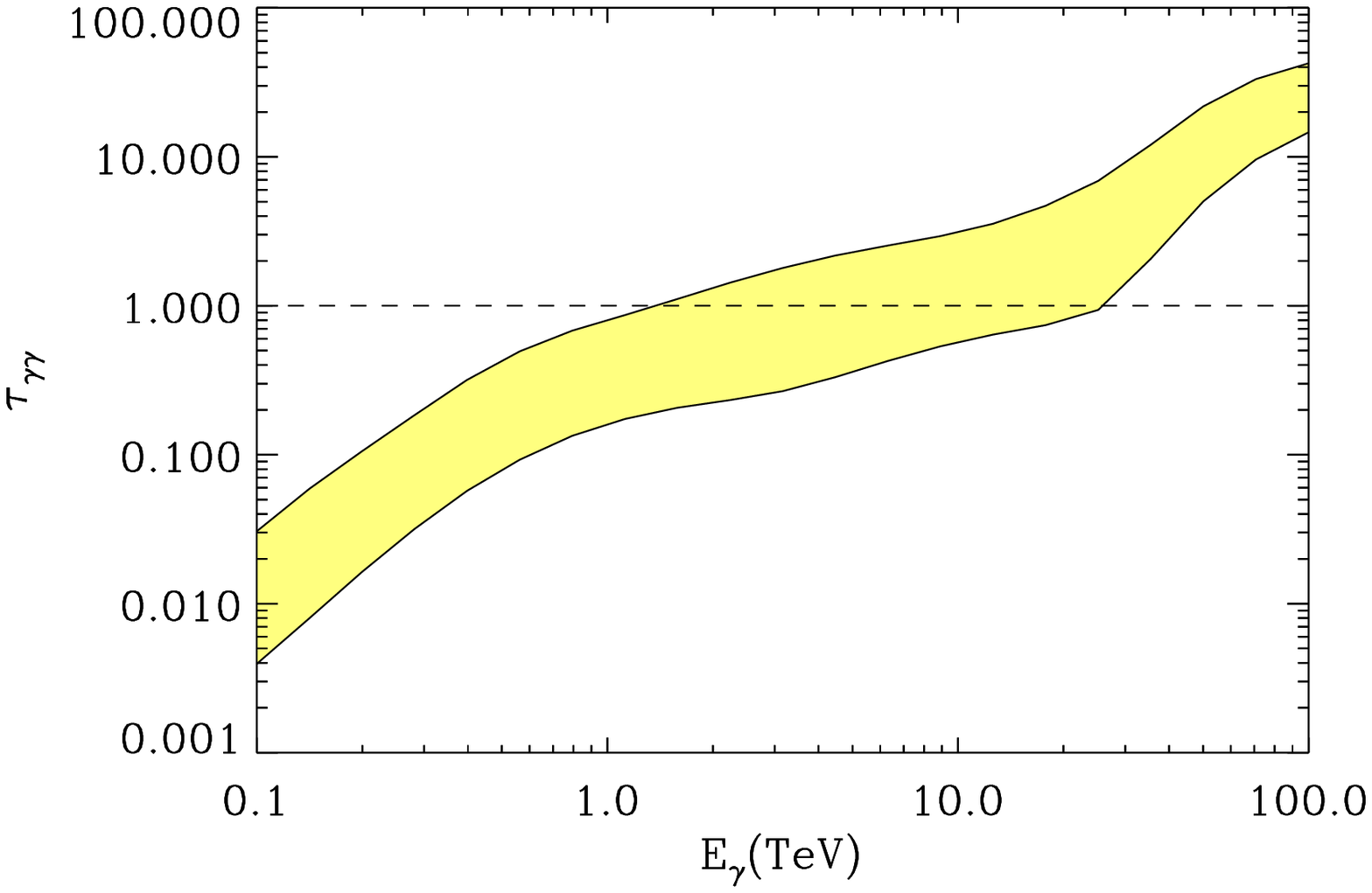}
\caption{}
\label{tau_plot}
\end{figure}

\begin{figure}
\epsfxsize=4.75in       
\epsfbox{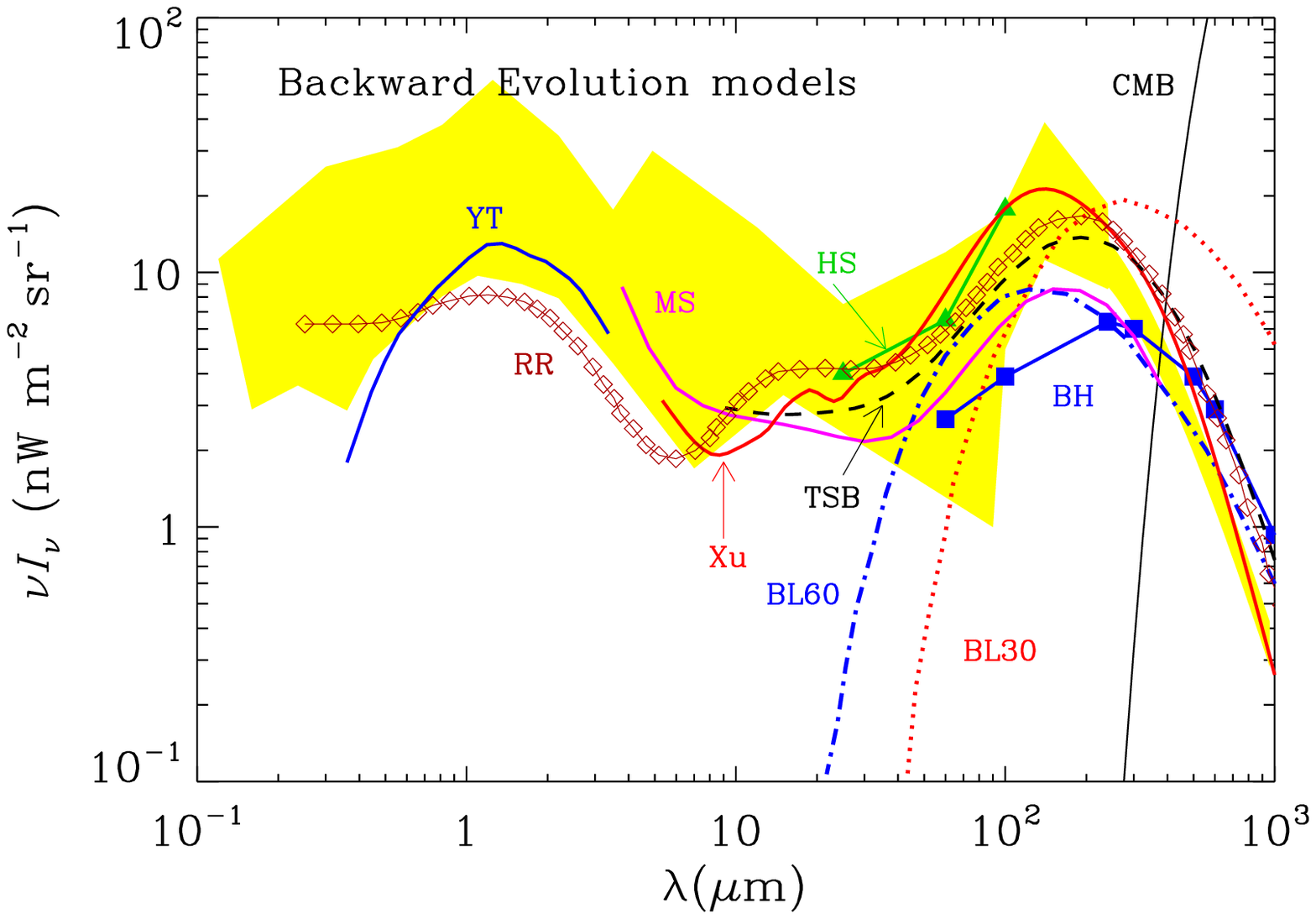}
\caption{}
\label{models_be}
\end{figure}

\begin{figure}
\epsfxsize=4.75in       
\epsfbox{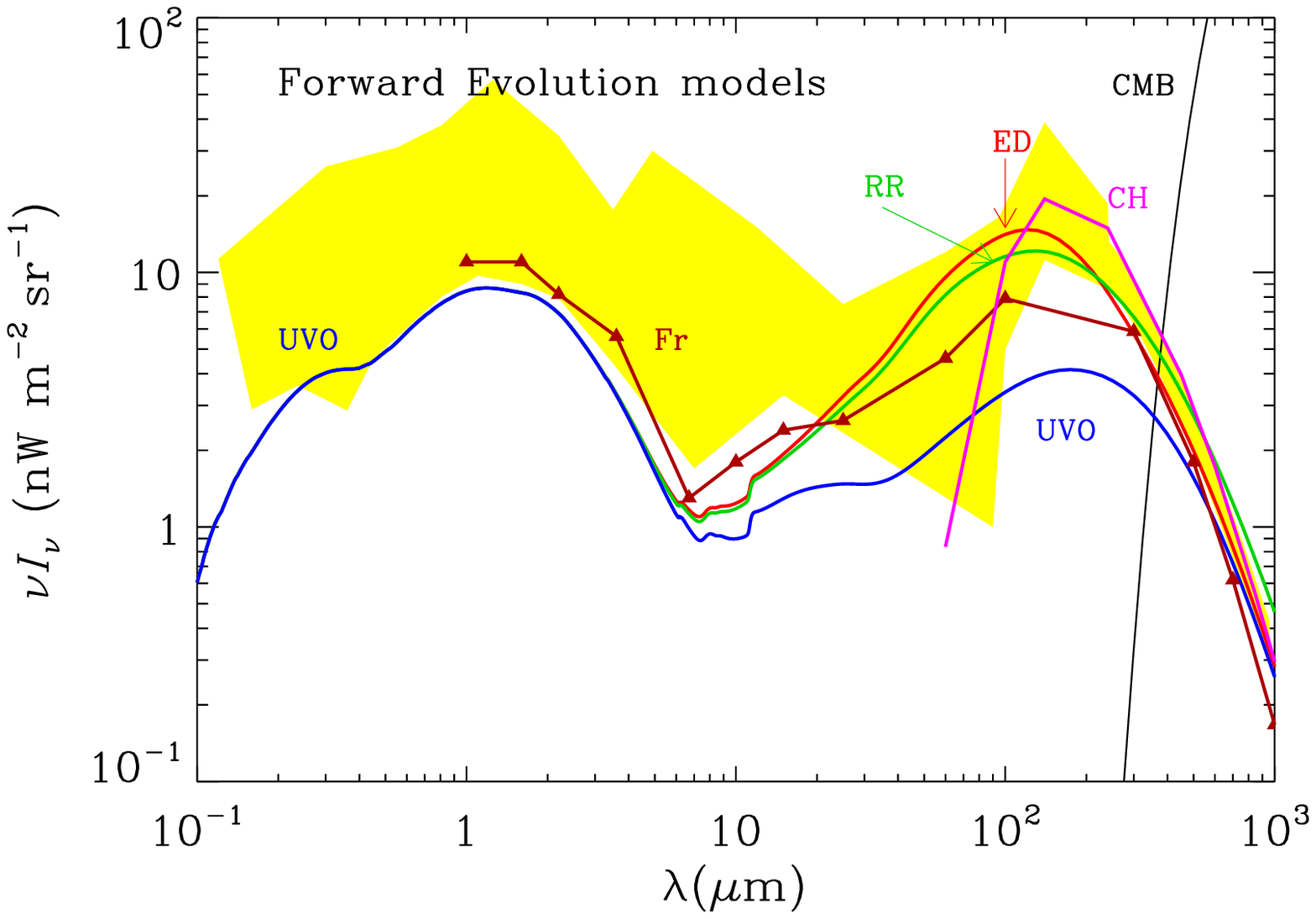}
\caption{}
\label{models_fe}
\end{figure}

\begin{figure}
\epsfxsize=4.75in       
\epsfbox{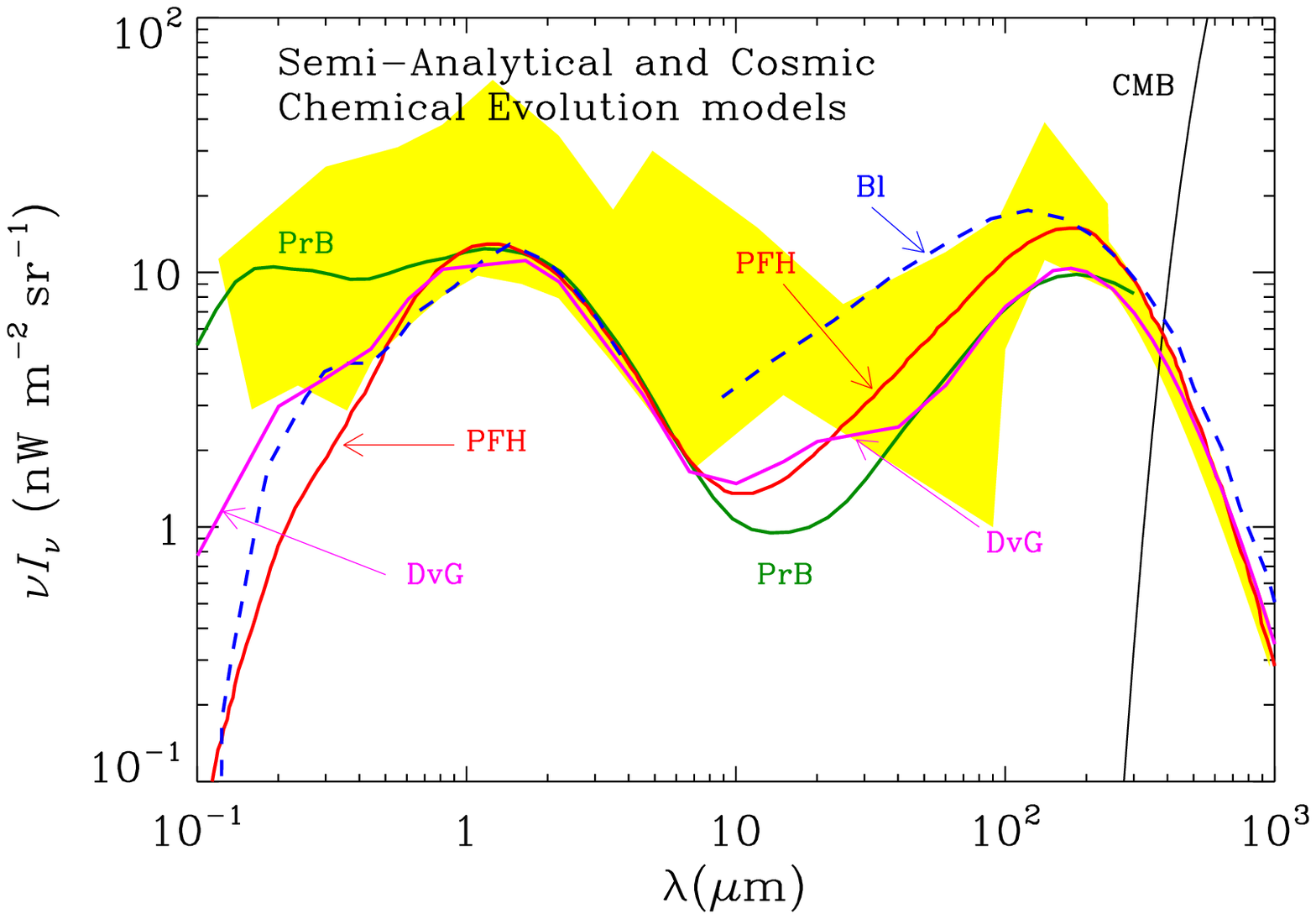}
\caption{}
\label{models_sacce}
\end{figure}

\begin{description}

\item[Figure 1.]  Spectrum of the cosmic background radiations.  The radio background (CRB) is represented by a $\nu I_{\nu} \propto \nu^{0.3}$ spectrum, normalized 
to the Bridle (1967) value at 170 cm. The cosmic microwave background (CMB) is represented by a blackbody spectrum at 2.725~K. The UV-optical (CUVOB) and infrared (CIB) backgrounds are schematic representations of the work summarized in this review (Figure~\ref{ebl_plot}).  The data for the X-ray background (CXB) are taken from Wu et al (1991), and the curves are analytical 
representations summarized by Fabian \& Barcons (1992). The $\gamma$-ray background (CGB) is represented by the 
power law given by Sreekumar et al (1998).

\item[Figure 2.]  Foreground contributions to the DIRBE data
  at \mbox{$1.25-240~\mu$m} in the Lockman Hole area: observed sky brightness
  (open circles), interplanetary dust (triangles), bright Galactic sources
  (squares), faint Galactic sources (asterisks), and the interstellar
  medium (diamonds).  Solid circles are the residuals after removing all foregrounds from the observed   brightness.  Adapted by permission from Hauser et al 1998.

\item[Figure 3.]  Diffuse extragalactic infrared background measurements and limits.  Error bars
shown are 1$\sigma$.  Discrete points are from DIRBE data (see references in Tables 1 and 2).  Curves at submillimeter wavelengths are from FIRAS data: blue, Fixsen et al (1998); red, Lagache et al (1999); green, Puget et al (1996).  The red curve in the near-infrared is from {\it IRTS} data, Matsumoto et al (2000).  

\item[Figure 4.]  Cosmic infrared background limits implied by TeV $\gamma$-ray observations. Dashed lines 
represent limits 
derived from Mrk 421 data: DJSS (De Jager et al 1994); DwSl (Dwek \& 
Slavin 1994); B98 (Biller et al 1998).   Solid lines represent limits derived from Mrk 
501 data: StF (Stanev \& Franceschini 1998); Funk (Funk et al 1998); Guy (Guy et al 2000).  Solid square limit is from Mrk 501 (Mannheim 1998). All limits were
scaled to a Hubble constant of H$_0$ = 100 km\, s$^{-1}$\, Mpc$^{-1}$ 
($\nu I_{\nu}\,\propto$~H$_0$).  The shaded region shows the extragalactic background light limits defined by the
data in Figure~\ref{ebl_plot}.

\item[Figure 5.]  Summary of extragalactic background light (EBL) measurements and limits.  Error bars for detections are 1$\sigma$.  Square symbols show lower limits obtained by integrating the light of detected sources.  X's show 2$\sigma$ lower limits on integrated resolved sources from Bernstein (1999).  Diamonds show upper limits from fluctuation measurements.  All other symbols show absolute background measurements (1$\sigma$ error bars) or limits (2$\sigma$).  The shaded region represents current observational limits for the EBL spectrum, and the dotted line shows nominal values (see \S~\ref{ebl_sum} for discussion).  The black line (CMB) shows the cosmic microwave background radiation.

\item[Figure 6.]  Limits for the TeV $\gamma$-ray optical depth, $\tau_{\gamma \gamma}$, as a function of $\gamma$-ray energy. The optical depth limits were derived from the EBL limits defined by the
shaded area in Figure~\ref{ebl_plot}. The optical depth was calculated for a source distance of $z$ = 0.03, and H$_0$ = 100 km\, s$^{-1}$\, Mpc$^{-1}$  ($\tau_{\gamma \gamma}\,\propto$~H$_0^{-1}$).

\item[Figure 7.]  Extragalactic background light (EBL) predicted by backward evolution models:  HS (Hacking \& Soifer 1991, solid triangles); BH (Beichman \& Helou 1991, solid squares); MS (Malkan \& Stecker 1998, purple line); BL30 (red dotted line), BL60 (blue dash-dot line) (Blain \& Longair 1993); Xu (Xu 2000, solid red line); TSB (Tan et al 1999, black dashed line); RR (Rowan-Robinson 2001, open diamonds); YT (Yoshii \& Takahara 1988, blue solid line).  The shaded region shows the EBL limits defined by the data in Figure~\ref{ebl_plot}. See details in \S~\ref{be}.

\item[Figure 8.]  Extragalactic background light (EBL) predicted by forward evolution models: Fr (Franceshini et al 1994, solid triangles); CH (Calzetti \& Heckman 1999, solid violet line); ED, RR, UVO (Dwek et al 1998; red, green, blue solid lines respectively).  The shaded region shows the EBL limits defined by the
data in Figure~\ref{ebl_plot}.  See details in \S~\ref{fe}.

\item[Figure 9.]  Extragalactic background light (EBL) predicted by models.  Semi-analytical: PrB (Primack et al 1999; green line); DvG (Devriendt \& Guiderdoni 2000, violet line); Bl (Blain et al 1999a, dashed blue line).  Cosmic chemical evolution: PFH (Pei et al 1999).  The shaded region shows the EBL limits defined by the 
data in Figure~\ref{ebl_plot}.  See details in \S~\ref{sa} and \S~\ref{cce}.

\end{description}

\newpage

\renewcommand\baselinestretch{2}

\begin{table}
\footnotesize
\caption{Infrared Extragalactic Background Measurements}
\label{cib_tab}
\begin{tabular}{@{}lllll@{}}
\hline\hline
\vspace{-0.1in}

 $\lambda$ &
 $\nu I_{\nu}$ &
 Isotropy &
 Claimed  &
 Reference \\

 ($\mu$m) &
 (nW m$^{-2}$ sr$^{-1}$) &
 test passed & detection
 &
 \\ \hline

\vspace{-0.15in}
  1.25  & $<$ 75 (33$\pm$21)         &  no      & no  & Hauser et al
1998 \\
\vspace{-0.15in}
  1.25   & $<$ 108 (60$\pm$24)    	 &  no	& no	& Dwek \& Arendt 1998 $^{(e)}$ \\
\vspace{-0.05in}
  1.25  & $<$ 57 (28$\pm$15)      &  no      & no  & Wright 2001b  \\

\vspace{-0.15in}
  2.2   & $<$ 39 (15$\pm$12)         &  no      & no  & Hauser et al
1998 \\
\vspace{-0.15in}
  2.2   &  23$\pm$6       &  yes  	& yes	& Wright \& Reese 2000 \\
\vspace{-0.15in}
 2.2   &   22$\pm$6        	 &  no      & tentative	& Gorjian et al 2000  \\
\vspace{-0.05in}
 2.2   &   20$\pm$5  	 	 &  yes     & yes	&  Wright 2001b \\

\vspace{-0.15in}
  3.5   & $<$ 23 (11$\pm$6)          &  no      & no  & Hauser et al
1998 \\
\vspace{-0.15in}
  3.5   & 14$\pm$3        	 &  no	& tentative	& Dwek \& Arendt 1998 $^{(e)}$ \\
\vspace{-0.15in}
  3.5   & 11$\pm$3        	 &  no	& tentative	&  Gorjian et al 2000 \\
\vspace{-0.05in}
  3.5   & 12$\pm$3         &  yes	& yes	&  Wright \& Reese 2000\\

\vspace{-0.15in}
  1.4-4 & (see Figure~3) &  yes	& yes	& Matsumoto et al 2000 \\

\vspace{-0.15in}
  4.9   & $<$ 41 (25$\pm$8)          &  no      & no  & Hauser et al
1998 \\
\vspace{-0.05in}
  4.9   & $<$ 38 (25$\pm$6)    	 &  no	& no	& Dwek \& Arendt 1998 $^{(e)}$ \\

\vspace{-0.05in}
  12    & $<$ 470 (190$\pm$140)      & no       & no  & Hauser et al
1998 \\
\vspace{-0.05in}
  25    & $<$ 500 (190$\pm$160)     & no       & no  & Hauser et al
1998 \\
\vspace{-0.15in}
  60    & $<$ 75 (21$\pm$27)         & no       & no  & Hauser et al
1998 \\

\vspace{-0.05in}
  60    & 28$\pm$7        	 &  yes	& tentative    & Finkbeiner et al 2000 \\
\vspace{-0.15in}
  100   & $<$ 34 (22$\pm$6)          & no       & no  & Hauser et al
1998 \\
\vspace{-0.15in}
  100   & $>$ 5  (11$\pm$3)          & no       & no  & Dwek et al 1998
\\
\vspace{-0.15in}
 100    & 23$\pm$6  	 &  no	& yes	& Lagache et al 2000 \\
\vspace{-0.05in}
 100    & 25$\pm$8        	 &  yes	& tentative	& Finkbeiner et al 2000 \\

\vspace{-0.15in}
  140   & 32$\pm$7          	 &  yes	& tentative	& Schlegel et al 1998 $^{(c)}$ \\
\vspace{-0.15in}
  140   & 25$\pm$7             & yes   & yes & Hauser et al
1998 $^{(c)}$ \\
\vspace{-0.15in}
  140   & 15$\pm$6         & yes   & yes & Hauser et al
1998 $^{(d)}$ \\
\vspace{-0.15in}
  140   & 15$\pm$6  	 	 & no		& yes   &  Lagache et al.
 1999 $^{(c)}$ \\
\vspace{-0.05in}
  140   & $<$ 47 (24$\pm$12)     &  no   & no    & Lagache et
al 2000 $^{(c)}$ \\

\vspace{-0.15in}
  240   & 17$\pm$2            	 &  yes	& tentative   & Schlegel et al 1998 $^{(c)}$ \\
\vspace{-0.15in}
  240   & 14$\pm$3         & yes  & yes & Hauser et al 1998 $^{(c)}$ \\
\vspace{-0.15in}
  240   & 13$\pm$2         & yes   & yes & Hauser et al 1998 $^{(d)}$ \\
\vspace{-0.15in}
  240   & 11$\pm$2  	 	 & no		& yes   &  Lagache et al. 1999 $^{(c)}$ \\
\vspace{-0.05in}
  240   & $<$ 25 (11$\pm$7)  	 & no		& no   &  Lagache et al. 2000 $^{(c)}$ \\

\vspace{-0.1in}
 200-1000 &  (see Figure~3) 
   & yes &  tentative & Puget et al. 1996 \\
\vspace{-0.1in}
 200-1000 & $a \left({\nu \over \nu_0}\right)^k \ \nu B_{\nu}(T)$
      & yes & yes & Fixsen et al 1998$^{(a)}$ \\
\vspace{0.1in}
 200-1000 & $a \left({\nu \over \nu_0}\right)^k \ \nu B_{\nu}(T)$
   & yes &  yes & Lagache et al 1999$^{(b)}$ \\ \hline

\end{tabular}

\vspace{0.1in}
Error bars are 1$\sigma$.  Upper and lower limits are at 95\% CL.
\vspace{-0.2in}
\newline
Values in parentheses are residuals and their uncertainties.
\vspace{-0.2in}
\newline
Column 3 indicates whether some degree of isotropy was demonstrated.
\vspace{-0.2in}
\newline
Column 4 indicates whether definite (yes), possible (tentative) or no detection was claimed. 
\vspace{-0.2in}
\newline
$^{(a)}\  a=(1.3\ \pm\ 0.4)\ \times\  10^{-5},\ k=0.64\ \pm\ 0.12,\ T=(18.5\ \pm\ 1.2)\ $K,\ 
$\lambda_0\ =\ 100\ \mu$m 
\vspace{-.2in}
\newline
$^{(b)}\  a=8.8\  \times\  10^{-5},\ k=1.4,\ T=13.6\  $K,\ $\lambda_0\ =\ 100\ \mu$m 
\vspace{-0.2in}
\newline
$^{(c)}$  Calibrated on DIRBE photometric scale 
\vspace{-0.2in}
\newline
$^{(d)}$  Calibrated on FIRAS photometric scale
\vspace{-0.2in}
\newline
$^{(e)}$  Based on CIB average $\nu$I$_{\nu}(2.2\ \mu$m) = 22$\pm$6~\nwat  
\newline
\normalsize
\end{table}


\begin{table}
\footnotesize
\caption{Measurements of Infrared Background Fluctuations}
\label{fluc_tab}
\begin{tabular}{@{}llllll@{}}
\hline\hline
\vspace{-.1in}
 $\lambda$&
 $\theta^{(a)}$&
 $\delta\,(\nu$I$_\nu$)~$^{(b)}$&
 P$_S$~$^{(c)}$&
 $\nu $I$_{\nu}$~$^{(d)}$&
Reference \\

 ($\mu$m)&
 (arcmin)&         
 (nWm$^{-2}$sr$^{-1}$)&
 (Jy$^2$ sr$^{-1}$)&
 (nWm$^{-2}$sr$^{-1}$)& 
  \\
\hline
\vspace{-0.1in}
1.25    &  42	& $< 19$   &  &  $< 200$	  &Kashlinsky et al 1996a \\
\vspace{-0.1in}
1.25    &  42	&  $15.5 {+3.7\atop -7.0}$	 &    	&  & Kashlinsky et al 2000$^{(e)}$ \\
\vspace{-0.1in}
1.25    &  42	& $<$ 8.2	       &    	&  & Wright 2001b \\
\vspace{-0.1in}
1.4	  &  8			& $\sim $18   &          & & Matsumoto et al 2000 \\
\vspace{-0.1in}
2.2     &  0.2--0.5	&  $< 9.6$    &    	& & Boughn et al 1986$^{(f)}$ \\
\vspace{-0.1in}
2.2     &  1--5		&  $< 4.1$    &    	& & Boughn et al 1986$^{(f)}$ \\
\vspace{-0.1in}
2.2     &  42	& $<$ 7	&       & $<$ 78  	& Kashlinsky et al 1996a \\
\vspace{-0.1in}
2.2    &  42	& $5.9 {+1.6\atop -3.7}$      &   &      & Kashlinsky et al 2000 $^{(e)}$ \\
\vspace{-0.1in}
2.2    &  42	& $<$ 4.0       &    &	& Wright 2001b \\
\vspace{-0.1in}
2.6    &   8	& $\sim $5     &	   &	& Matsumoto et al 2000 \\
\vspace{-0.1in}
3.5     &  42	& $< 2.4$	 &  &   $<$ 26  	& Kashlinsky et al 1996a \\
\vspace{-0.1in}
3.5    &  42	& $2.4 {+0.5\atop -0.9}$       &    &	& Kashlinsky et al 2000 $^{(e)}$ \\

\vspace{-0.1in}
4.9    &  42	& $< 1.3$   &  &  $< 13$	  &Kashlinsky et al 1996b \\
\vspace{-0.1in}
4.9    &  42	& $2.0 {+0.25 \atop -0.5}$     &    &	& Kashlinsky et al 2000 $^{(e)}$ \\
\vspace{-0.1in}
12--100 &  42	& $\lesssim $1--1.5     &  & $\lesssim $10--15 	& Kashlinsky et al 1996b \\
\vspace{-0.1in}
12      &  42	& $<$1.0     &      & $<$15	& Kashlinsky et al 2000 \\
\vspace{-0.1in}
25      &  42	& $<$0.5     &       & $<$8	& Kashlinsky et al 2000 \\
\vspace{-0.1in}
60      &  42	& $<$0.8     &      & $<$12	& Kashlinsky et al 2000 \\
\vspace{-0.1in}
100      &  42	& $<$1.1     &      &  $<$17	& Kashlinsky et al 2000 \\

\vspace{-0.1in}
90	  &  0.4--20      &  &$13000\pm3000$ &    $> 3$--10    & Matsuhara et al 2000  \\
	  &			&  &	(150 mJy)     &                 &                         \\

\vspace{-0.1in}
170	  &  0.6--4       &  $\sim\ $1 &  7400		      &   & Lagache \& Puget 2000 \\
	  &			&  &	(100 mJy)     &                 &                         \\
\vspace{-0.1in}
170	  &  0.6--20      &  &  $12000\pm2000$ & $>0.9$--2.6   & Matsuhara et al 2000 \\
	  &			&  &	(250 mJy)     &                 &                         \\

400--1000 &  420 & $4(\lambda /$400 $\mu$m)$^{-3.1}$ &   & & Burigana \& Popa 1998 \\ \hline
\end{tabular}

\vspace{0.12in}
$^{(a)}$ Approximate angular scale of fluctuation measurements. 
\newline
$^{(b)}$  [$\delta\,(\nu$I$_\nu$)]$^2$ is the variance of $\nu $I$_\nu$.  
\newline
$^{(c)}$ Source power spectrum, for sources fainter than flux shown in parentheses. 
\newline
$^{(d)}$  Reported limit on the CIB inferred from the fluctuation measurement.
\newline
$^{(e)}$  Limits give 92\% confidence interval.
\newline
$^{(f)}$  90\% confidence level.
\newline
\normalsize
\vspace{-.2in}
\end{table}


\newpage
\begin{table}
\small
\caption{Integrated galaxy light$^{(a)}$}
\label{count_tab}
\begin{tabular}{@{}llll@{}}
\hline\hline
  $\lambda(\mu$m) & $\nu I_{\nu}$(nW m$^{-2}$ sr$^{-1}$) &  Comments
& Reference \\

\hline

\vspace{-0.08in}
   0.1595  & {\bf $>$2.9${+0.6\atop -0.4}$} &{\it HST}/STIS	    &  Gardner et al 2000 \\	
\vspace{-0.08in}
   0.1595  & {\bf $<$3.9${+1.1\atop  -0.8}$}  &{\it HST}/STIS	    &  Gardner et al 2000 \\
\vspace{-0.08in}
   0.2     & 0.6                          & FOCA   &  Milliard et al 1992 \\
\vspace{-0.08in}
   0.2365  &  {\bf 3.6${+0.7\atop -0.5}$}	& {\it HST}/STIS      &  Gardner et al 2000 \\	
\vspace{-0.08in}
   0.300  &  $>$9        &   {\it HST}/WFPC2 & Bernstein 1999 \\
\vspace{-0.08in}
   0.36  &  {\bf 2.9${+0.6\atop -0.4}$}   & (see Ref.)   & Madau \& Pozzetti 2000 \\
\vspace{-0.08in}
   0.45  &  {\bf 4.6${+0.7\atop -0.5}$}   & (see Ref.)   & Madau \& Pozzetti 2000 \\
\vspace{-0.08in}
   0.555  &  $>$8        &   {\it HST}/WFPC2 & Bernstein 1999 \\
\vspace{-0.08in}
   0.67  &  {\bf 6.7${+1.3\atop -0.9}$}   & (see Ref.)   & Madau \& Pozzetti 2000 \\

\vspace{-0.08in}
   0.81  &  {\bf 8.0${+1.6\atop -0.9}$}   & (see Ref.)   & Madau \& Pozzetti 2000 \\
\vspace{-0.08in}
   0.814  & $>$12        &   {\it HST}/WFPC2 & Bernstein 1999 \\
\vspace{-0.08in}
   1.1  &  {\bf 9.7${+3.0\atop -1.9}$}   &  {\it HST}/NICMOS  & Madau \& Pozzetti 2000 \\
\vspace{-0.08in}
   1.6  &  {\bf 9.0${+2.6\atop -1.7}$}   &  {\it HST}/NICMOS  & Madau \& Pozzetti 2000 \\
\vspace{-0.08in}
   2.2  &  {\bf 7.9${+2.0\atop -1.2}$}   &  {\it HST}/NICMOS  & Madau \& Pozzetti 2000 \\

\vspace{-0.08in}
    7     & 1.7$\pm$0.5       &  {\it ISO}/ISOCAM      &  Altieri et al 1999 \\
\vspace{-0.08in}
   12    &  0.50$\pm$0.15 &  {\it ISO}/ISOCAM  & Clements et al 1999 \\
\vspace{-0.08in}
   15    &  3.3$\pm$1.3   &  {\it ISO}/ISOCAM  & Altieri et al 1999 \\
\vspace{-0.08in}
   25    &  0.02      &  {\it IRAS}  & Hacking \& Soifer 1991 \\
\vspace{-0.08in}
   60    &  0.4       & {\it IRAS}  & Hacking \& Soifer 1991 \\

\vspace{-0.08in}
   90   & 1.0	     &  {\it ISO}/ISOPHOT  &  Matsuhara et al 2000 \\
\vspace{-0.08in}
   90   & $\sim$0.9	     &  {\it ISO}/ISOPHOT  &  Juvela et al 2000 \\
\vspace{-0.08in}
   95   & 0.5          &  {\it ISO}/ISOPHOT  &  Kawara et al 1998 \\
\vspace{-0.08in}
  100   &   0.2       &  {\it IRAS}  & Hacking \& Soifer 1991 \\

\vspace{-0.08in}
  150   & $\sim$1.0          &  {\it ISO}/ISOPHOT  &  Juvela et al 2000 \\
\vspace{-0.08in}
  170   & 0.35          &  {\it ISO}/ISOPHOT  &  Kawara et al 1998 \\
\vspace{-0.08in}
  170   & 0.9          &  {\it ISO}/ISOPHOT  &  Matsuhara et al 2000 \\
\vspace{-0.08in}
  175   & 1.75          &  {\it ISO}/ISOPHOT  &  Puget et al 1999 \\
\vspace{-0.08in}
  180   & $\sim$1.2          &  {\it ISO}/ISOPHOT  &  Juvela et al 2000 \\

\vspace{-0.08in}
  850   & 0.11    &   SCUBA  & Barger et al 1999a \\
  850	  & 0.5$\pm$0.2     &   SCUBA  & Blain et al 1999b \\
\hline
\end{tabular}

$^{(a)}$ {\bf Bold face} indicates counts to a depth where the light has converged.  Inequalities indicate limits on the integrated galaxy light, not the EBL.  All entries are regarded as lower limits to the EBL.  
\newline

\normalsize

\end{table}


\begin{table}
\footnotesize
\caption{UV-optical extragalactic background light}
\label{uvo_tab}
\begin{tabular}{@{}llll@{}}
\hline\hline
\vspace{-0.1in}

 $\lambda$       &      $\nu I_{\nu}$    & Comments       &       Reference \\
 ($\mu$m)        &      (nW m$^{-2}$ sr$^{-1}$) &        &                \\

\hline
0.10     & $<$ 1.2 (0$\pm$0.6)	&  Voyager UVS & Murthy et al 1999 \\
0.10	   & $<$ 11				&  Voyager UVS & Edelstein et al 2000 \\
0.1595   & $<$ 14 (10$\pm$2)		&  HST/STIS    & Brown et al 2000 \\
0.165    & $<$ 7.0 (5.6$\pm$0.7)	&  Shuttle UVX  & Martin et al 1991 \\
0.300    & 12$\pm$7       		&  HST \& LCO   & Bernstein 1999 \\
0.4000   & 37$\pm$11			& ``Dark cloud" method & Mattila 1976 \\
0.4000   & $<$ 46 (26$\pm$10)		& ``Dark cloud" method  & Mattila 1990  \\
0.41     & $<$ 10                   &  OAO-2/WEP   & Lillie 1968 \\
0.4400   & $<$ 20 (7$\pm$7)		&  Pioneer 10  & Toller 1983 \\
0.4400   & $<$ 60 (10$\pm$25)		&  Pioneer 10  &  Toller 1983$^{(a)}$  \\
0.44     & $<$ 26 (1$\sigma$)       & ``Dark cloud" method  & Spinrad \& Stone 1978 \\
0.5115   & $<$26 (8$\pm$9)		&  Ground-based photometer & Dube et al  1979 \\
0.5115   & $<$48 (30$\pm$9)		&  Ground-based photometer & Dube et al  1979$^{(a)}$  \\
0.53     & $<$ 33  			&  Ground-based photometer  & Roach \& Smith 1968 \\
0.555    & 17$\pm$7       		&  HST \& LCO    & Bernstein 1999 \\
0.814    & 24$\pm$7       		&  HST \& LCO    & Bernstein 1999 \\ \hline

\end{tabular}
\vspace{0.1in}
\newline
Error bars are 1$\sigma$.  Upper limits are 2$\sigma$ unless otherwise noted.
\vspace{-0.1in}
\newline
Values in parentheses are measurements and their 1$\sigma$ uncertainties.
\newline
$^{(a)}$  Value revised according to Leinert et al (1998).

\normalsize
\end{table}

\end{document}